\documentclass[%
 aip,
 jmp,%
 amsmath,amssymb,
 reprint,%
]{revtex4-1}

\usepackage{graphicx}
\usepackage{dcolumn}
\usepackage{bm}
\usepackage{xcolor}

\begin{document}

\preprint{AIP/123-QED}

\title{A ternary map of Ni-Mn-Ga Heusler alloys from \textit{ab initio} calculations}

\author{Yu. Sokolovskaya}%
 
\affiliation{ 
Faculty of Physics, Chelyabinsk State
University, Chelyabinsk, 454001, Russia}%
\author{O. Miroshkina}
 \affiliation{
Faculty of Physics, Chelyabinsk State
University, Chelyabinsk, 454001, Russia}%
\author{V. Sokolovskiy}%
\affiliation{ 
Faculty of Physics, Chelyabinsk State
University, Chelyabinsk, 454001, Russia}%
\author{M. Zagrebin}
\email{miczag@mail.ru}
 \affiliation{
Faculty of Physics, Chelyabinsk State
University, Chelyabinsk, 454001, Russia}%
 \affiliation{
South Ural State University (national research university), Chelyabinsk, 454080, Russia}%

\author{V. Buchelnikov}
 \affiliation{
Faculty of Physics, Chelyabinsk State
University, Chelyabinsk, 454001, Russia}%

\author{A. Zayak}
 \affiliation{
Bowling Green State University, Bowling Green, OH 43403, USA}%

\date{\today}

\begin{abstract}
In the present work, the aspects of magnetic and structural properties of Ni-Mn-Ga alloys are described in the framework of first-principles approach and mapped into ternary composition diagrams. The stable atomic arrangement and magnetic alignment for compositions with cubic austenite and tetragonal martensite structures across phase diagrams are predicted. It is shown that Ni- and Ga-rich compositions possess the regular Heusler structure in contrast to Mn-rich compositions with inverse Heusler structure as favorable one. Compositions with unstable austenite structure are concentrated in the left and right sides of diagram whereas compositions with unstable martensite structure are located in the low-middle part of diagram. The magnetic phase diagrams showing regions with the ferromagnetic order and the complex ferrimagnetic order for austenitic and martensitic compositions are obtained. The results of calculations are in a good agreement with available experimental data.

\end{abstract}  
\maketitle
\section{Introduction}

During last decades, shape memory Ni-Mn-based Heusler alloys have attracted tremendous attention as promising materials for a wide variety of engineering applications in the areas of magnetic actuators and sensors, damping, magnetic cooling technology and other. One of the prototype of Heusler alloys is Ni$_2$MnGa, which has been studied intensively for more than 30 years. The distinguishing feature of Ni$_2$MnGa is a martensitic transformation in ferromagnetic (FM) state between the high- temperature austenitic phase with L2$_1$-cubic structure  ($Fm\bar3m$, space group no. 225) and a low-temperature martensitic phase with a modulated twin structure~\cite{Webster-1984} at $T_m$ about 200 K. The idea of magnetically induced martensitic variant reorientation in Ni$_2$MnGa has been suggested by Ulakko et al.~\cite{Ullakko-1996}. They were the first to demonstrate magnetically induced strains of 0.2\% along the [001] direction in a Ni$_2$MnGa single crystal at a temperature of 265 K and a magnetic field of 0.8 T. That study laid the foundation for an intensive experimental and fundamental studies of different compositions of Ni-Mn-Ga alloys to cover various aspects, such as the crystal structure of the austenite and martensite~\cite{Chernenko-1998,Wedel-1999, Glavatska-2003,Gavriljuk-2003,Vasilev-2003}, magnetic and magnetoresistance properties~\cite{Dubenko-2012,Dubenko-2015}, thermally and magnetically induced deformation~\cite{Ullakko-1996, Sozinov-2002,Buche-2006}, magnetocaloric properties~\cite{Planes-2009,Kamantsev-2015}, heat treatment processes~\cite{Chernenko-1998,Pons-2000}, phase diagrams~\cite{Bozhko-1998,Vasilev-1999, Khovaylo-2005, Entel-2006,Jin-2002,buche-2007} and other. 

At present, the most studied compositions of Ni-Mn-Ga alloys are concentrated in a relatively small area located near the stoichiometric Ni$_2$MnGa despite the fact that the functional properties as well as magnetic and martensitic transition temperatures are very sensitive to compositions or their valence electron concentration $e/a$. 
For instance, a well-accepted and best-studied samples are Ni$_{2+x}$Mn$_{1-x}$Ga in the composition range $0 \leq x < 0.4$ ($7.5 \leq e/a < 7.8$)~\cite{Khovaylo-2005} and Ni$_2$Mn$_{1+x}$Ga$_{1-x}$ in the composition range $0 \leq x < 0.6$ ($7.5 \leq e/a < 8.0$)~\cite{Cakir-2013}. For some of those compositions, the coupled magnetostructural phase transition can occur. 
However, other stoichiometric Mn$_2$NiGa and Ga$_2$MnNi and their derivatives among the Ni-Mn-Ga family have received less consideration. Nevertheless, experimental findings~\cite{Liu-2005,Liu-2006,Singh-2010}
 have suggested that 
 Mn$_2$NiGa crystalizes to the inverse Heusler structure ($F\bar43m$, space group no. 216) and undergoes the martensitic transformation in the ferrimagnetic (FIM) state at $T_m$ = 270~K with 
a large thermal hysteresis up to 50 K. In addition, the highest Curie temperature $T_C$ of 588 K was observed. The inverse Heusler structure in Mn$_2$NiGa has been confirmed theoretically by \textit{ab initio} methods~\cite{Barman-2007,Chakrabarti-2013}. 

It is apparent that the deviation from stoichiometry can enhance or reduce effects observed and predict new features of compounds. Often, the complexity of such studies can be caused by the problem of the synthesis of non-stoichiometric compositions, as well as their mechanical stability. On the other hand, the predictive nature of first-principle methods allows to study theoretically a wide variety of compositions beyond their stoichiometry. The most theoretical investigations of Ni-Mn-Ga have concentrated on the properties of Ni$_{2+x}$Mn$_{1-x}$Ga and Ni$_2$Mn$_{1+x}$Ga$_{1-x}$~(for instance see Refs.~\cite{Uijttewaal-2009,Buchelnikov-2010,Li-2011,Hickel-2012,Xu-2015,Dutta-2016}). There were few attempts performed by our group to mapping structural and magnetic properties onto a ternary diagram for Ni-Mn-Ga in austenitic phase~\cite{Sokolovskaya-2017,Sokolovskiy-2019a,EPJ-2018}. The total energy calculations have been done assuming the electronic minimization scheme only, which is the indirect evidence of structure stability.

In the present work, using \textit{ab initio} density functional theory and
full geometry optimization,
we focus on the crystal structure and magnetic properties of austenitic and martensitic Ni-Mn-Ga in a wide compositional range that completely covers a ternary diagram.

\begin{figure*}[t!!!] 
\centerline{
\includegraphics[width=12.0cm,clip]{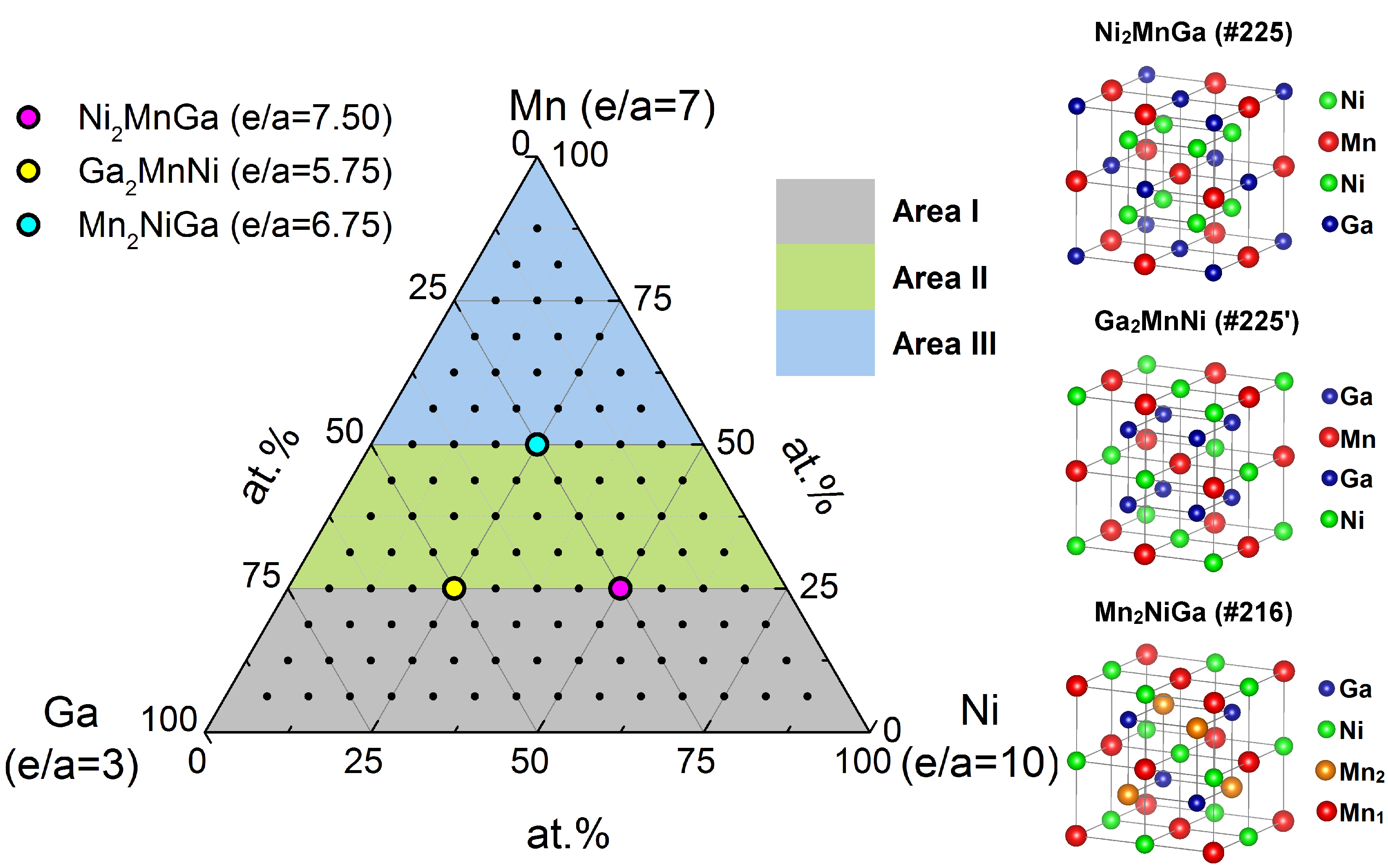}
           }
\caption{(Color online)
Ni-Mn-Ga compositions and crystal cubic structures considered in the \textit{ab initio} calculations assuming 16-atom supercell. Color symbols denote the stoichiometric Ni$_{50}$Mn$_{25}$Ga$_{25}$, Mn$_{50}$Ni$_{25}$Ga$_{25}$, and Ga$_{50}$Mn$_{25}$Ni$_{25}$. Here the area I includes compositions with Mn content $x \leq 4$ atoms ($x \leq 25$ at.\%); the area II contains compositions with Mn content $4 <x < 8$ atoms ($25 < x < 50$ at.\%); the area III indicates compositions with Mn content $x \geq 8$ atoms ($\geq 50$ at.\%). }
\label{Fig-1}
\end{figure*}

\section{Computational methodology}
The electronic structure calculations of the crystal structure stability as well as magnetic and structural properties are performed using the the projector augmented wave (PAW) method implemented in the Vienna \textit{ab initio} simulation package VASP~\cite{Kresse-1996, Kresse-1999}.  The exchange correlation effects are treated in the Perdew-Burke-Ernzerhof parametrization of 
generalized gradient approximation~\cite{Perdew-1996}. The PAW pseudopotentials for Ni(3$p^6$3$d^8$4$s^2$), Mn(3$p^6$3$d^6$4$s^1$), Ga(3$d^{10}$4$s^2$4$p^1$) are considered. 
The uniform Monkhorst-Pack $k$-point mesh ($8\times8\times8$) is used to sample the Brillouin zone.
The crystal structures of austenitic and martensitic phase as well as 
both lattice parameters and atomic coordinates are optimized 
using the conjugate-gradient algorithm for the minimization of forces acting on the atoms. 
The convergence criterion for the residual forces is $10^{-5}$eV/\AA~as well as all calculations are converged to an accuracy of $10^{-6}$ eV. 

To form various compositions covered fully the ternary diagram of Ni-Mn-Ga, 16-atom supercell is taken into consideration depending on a content. Thus, 105 compositions are generated with the concentration step of 6.25~at.\% as shown in Fig.~\ref{Fig-1}. Three of them are stoichiometric compositions Ni$_2$MnGa (Ni$_8$Mn$4$Ga$_4$), Mn$_2$NiGa (Mn$_8$Ni$_4$Ga$_4$), and Ga$_2$MnNi (Ga$_8$Mn$_4$Ni$_4$). In dependence of Mn content $x$, the ternary diagram can be splitted into three areas, which are labeled as follows: area I ($x \leq 25$ at.\%), area II ($25 < x < 50$ at.\%), and area III ($x \geq 50$ at.\%).
For Ni- and Ga-excess compositions (areas I and II), two regular Heusler structures with $Fm\bar3m$ symmetry (space group no. 225) are assumed. The main difference between them is a occupation of Ni and Ga at 8$c$-Wyckoff's sites, while 4$b$ site is occupied by Mn atoms. For Mn-excess compositions (area III), the inverse Heusler structure with $F\bar43m$ symmetry (space group no. 216), in which 
 Mn atoms are placed at 4$a$ and 4$c$ Wyckoff sites, is considered.
 Positions of Ni, Mn and Ga atoms for all considered types of crystal lattices are shown in the Supplemental Materials (SM)~\cite{SM}.
 
 Since the excess Mn atoms placed at Ni or Ga sites can interact antiferromagnetically with Mn atoms located at regular sites, the various magnetic configurations should be assumed.  For area I, the ferromagnetic (FM) spin order is considered only, while in the case area II, FM state and 
  two ferrimagnetic (FIM) states labeled as FIM-1 and FIM-2 are taken into account. For FIM-1 and FIM-2, the Mn excess atoms located at Ni and Ga sites have a reversed magnetic moment to that for Mn atoms, which occupy the regular sites. Finally, for area III, FM state and seven FIM states are accounted due to a various combinations of Mn magnetic moments. The list of all magnetic configurations is presented in SM~\cite{SM}. 
 
 To analyze the stability for 105 calculated compounds in austenite and martensite phase, a two-step procedure is used. In the first step, the 
geometric optimization calculations for compositions with initial cubic and tetragonal structure are carried out. The guess lattice parameters are taken from our previous calculations assuming electronic relaxation only~\cite{EPJ-2018}. In the second step, 
a minimal criterion in terms of formation energy is proposed for energetically favorable compositions. It is assumed that a structure will be stable if the total energy of Ni-Mn-Ga structure is less than the sum of total energies of its elements, namely $\Delta E_{form} =E_0^{\mathrm{Ni}-\mathrm{Mn}-\mathrm{Ga}} - (xE_0^{\mathrm{Ni}} + yE_0^{\mathrm{Mn}} + zE_0^{\mathrm{Ga}}) < 0$, here $x$, $y$, and $z$ are the atomic percents of Ni, Mn, and Ga.

\section{Results and Discussion}
\begin{figure}[b] 
\centerline{
\includegraphics[width=10.0cm,clip]{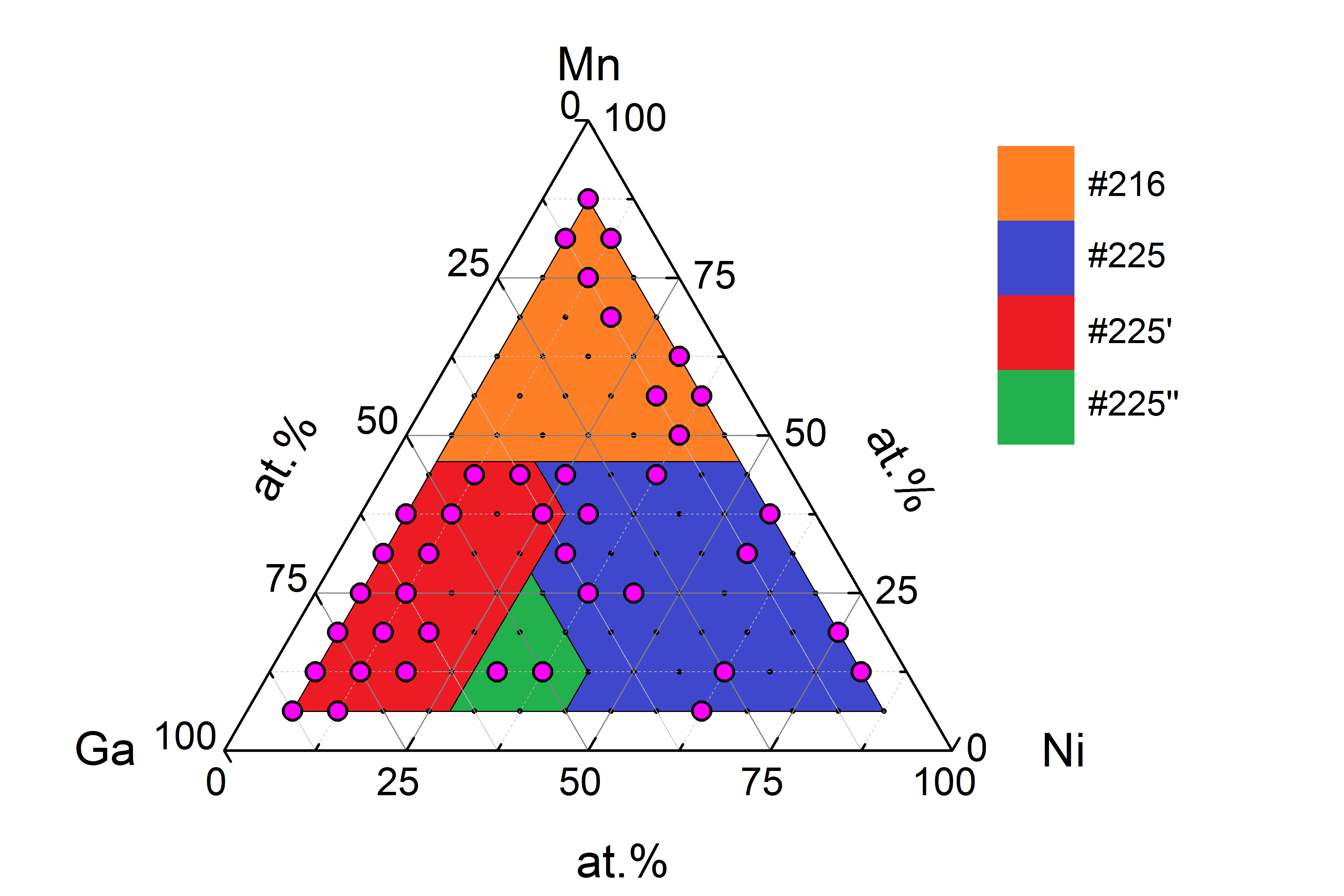}
           }
\caption{\label{fig:epsart} (Color online)
Ternary phase diagram in austenitic phase of Ni-Mn-Ga alloys. Circle symbols indicate the  unstable compositions, which have been relaxed to the distorted structures of low-symmetry martensite. The label \#216 denotes the inverse Heusler structure, while the labels \#225, \#225$^{\prime}$, and \#225$^{\prime\prime}$ indicate the regular Heusler structure with an occupation of 8$c$ Wyckoff sites as follows: for \#225, Ni atoms occupy 8$c$ sites; for\#225$^{\prime}$, Ga atoms are placed at 8$c$ sites; for \#225$^{\prime\prime}$, Ga and Ni atoms are located at 8$c$ sites.
}
\label{Fig-2}
\end{figure}

In the following, we discuss firstly the stability of austenitic phase in compounds considered. Figure~\ref{Fig-2} presents the structural phase diagram of austenite obtained from 
the ionic relaxation calculations.
We find that 65 compositions possess a cubic structure with a favorable atomic arrangement as follows: (i)
 the compounds with Mn content of $y \geq 50$~at.\% have the inverse Heusler structure ($\#$ 216); (ii) the compounds with Ni content of $x \geq 45$~at.\% and Mn content in the range of 
$25 < y < 50$~at.\% possess the regular Heusler structure ($\#$ 225), in which Ni atoms occupy 8$c$ sites; (iii)
the compounds with Ni and Mn content of $x < 45$~at.\% and $y < 25$~at.\%
show the regular Heusler structure ($\#$ 225), in which Ga atoms placed at 8$c$ sites.
In passing it should be mentioned that 40 compounds with initial cubic structure 
became tetragonally and orthorombically distorted structures at the end of relaxation. These are 18 compositions with Ga and Ni excess more than 60 at.\% and compositions with almost equiatomic Ni and Ga content (area I), 13 compositions with almost equiatomic Ni and Mn as well as Mn and Ga atoms (area II), and 9 compositions with Mn excess more than 50 at.\% (area III). The geometric optimization carried out for compounds with the initial tetragonal structure has revealed that 13 compositions are found to be energetically unfavorable due to higher total energy as compared to austenitic phase. These are 11 Ga-rich and almost equiatomic Ni and Ga compositions from the area I, and two compositions with almost equal Mn and Ga content from the area II. The list of stable compositions is presented in SM~\cite{SM}. We would like to emphasize
the quantity of unstable compounds in austenite phase resulting from ionic relaxation calculations
 is found to be larger as compared to those obtained from electronic relaxation calculations~\cite{EPJ-2018}.

\begin{figure}[t] 
\centerline{
\includegraphics[width=10.0cm,clip]{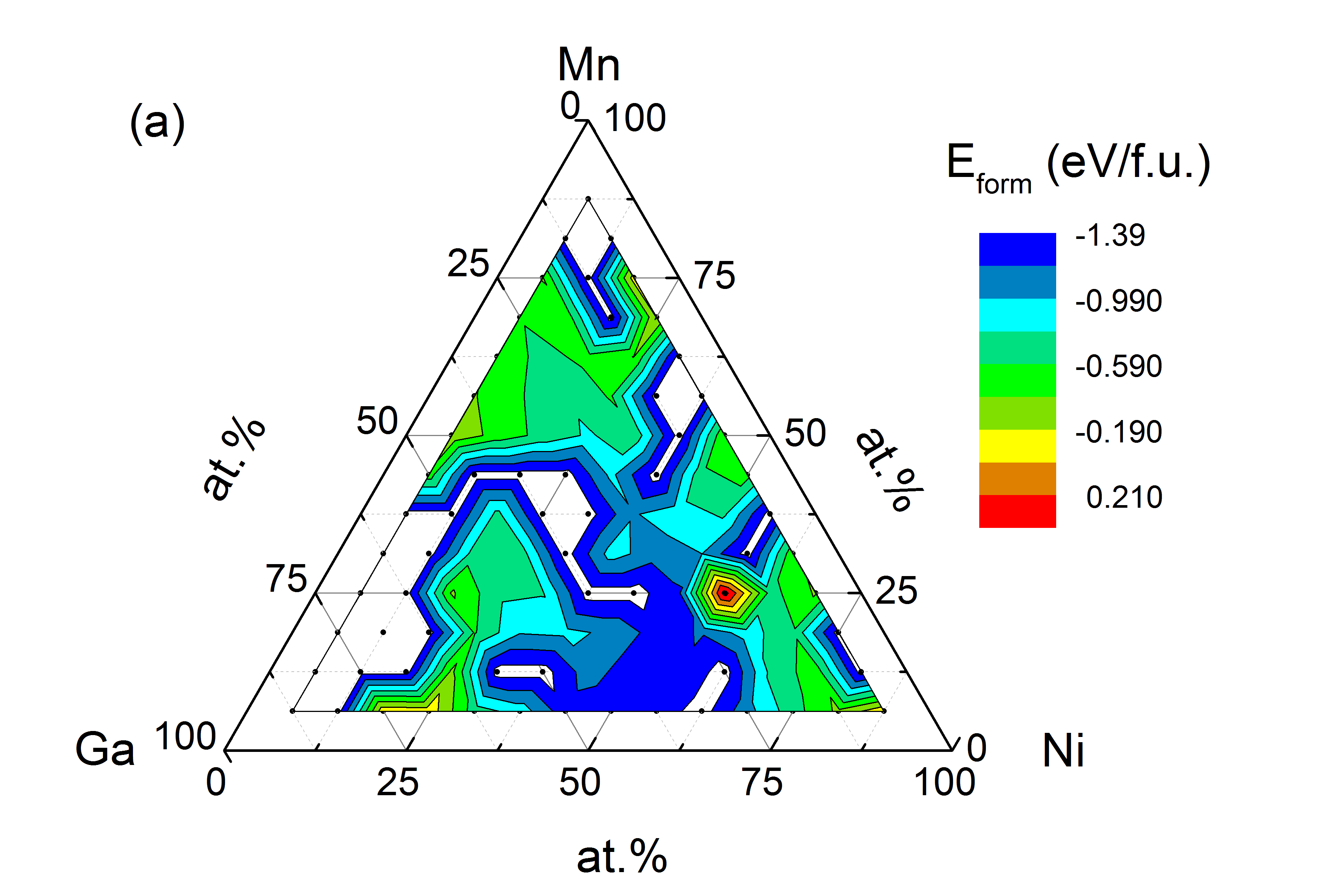}
           }
           \vspace*{0.3cm}
           \centerline{
\includegraphics[width=10.0cm,clip]{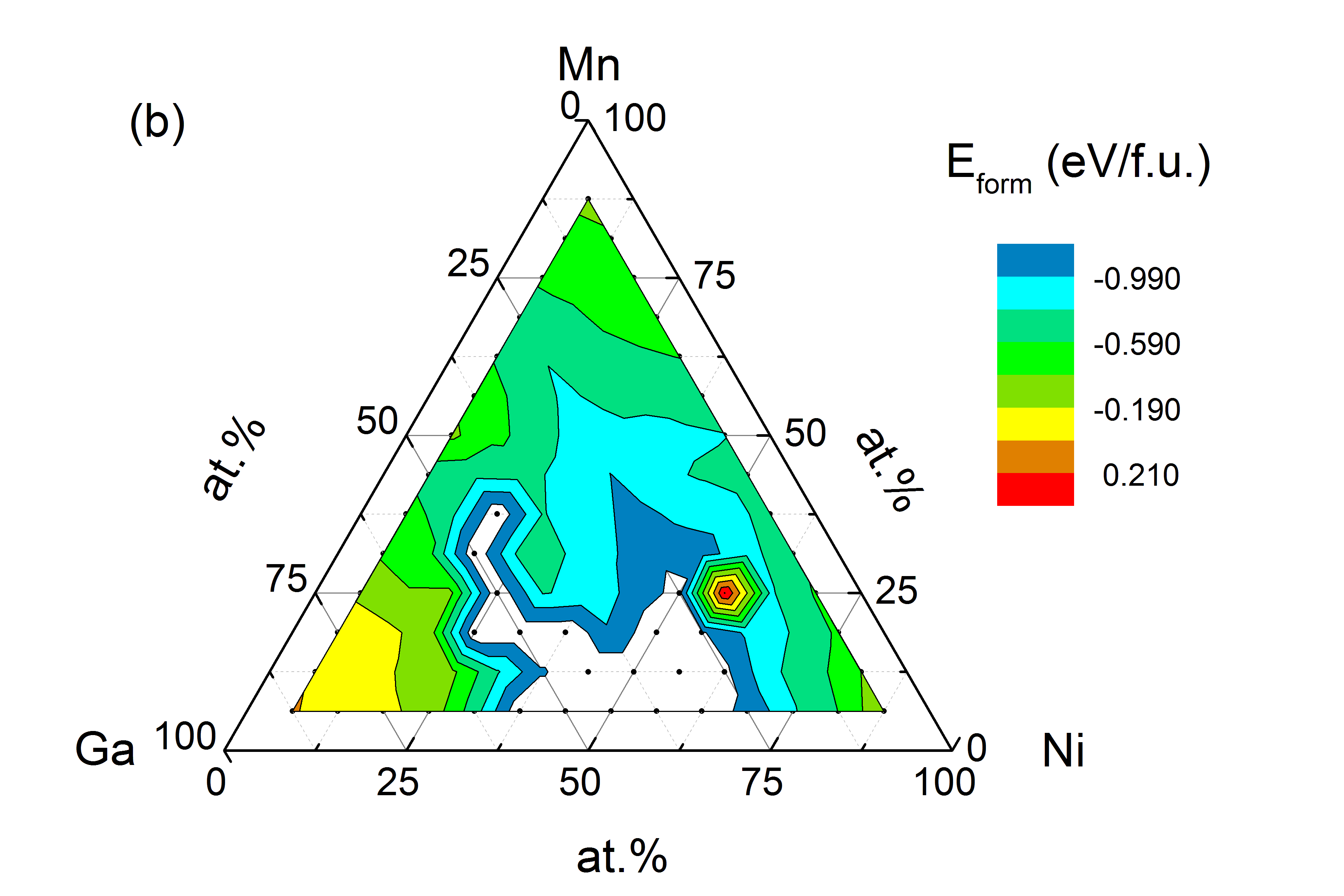}
           }
\caption{\label{fig:epsart} (Color online)
Formation energy mapped into the ternary diagram for Ni-Mn-Ga compositions in (a) austenite and (b) martensite.  White-colored regions indicate the unstable compositions in austenitic and martensitic phases.
}
\label{Fig-3}
\end{figure}

In Fig.~\ref{Fig-3} we show the formation energy for compositions with austenite and martensite phase mapped onto the ternary diagram. As can be seen the formation energy is negative for all compositions depicted indicating their possible stability. Compositions located at the middle part of diagram have larger negative values of the formation energy in comparison with Ga-, Ni-, and Mn-rich compositions.  Nevertheless, these observations suggest that compositions with a small formation energy should be stable at high temperatures.
  It is worth mentioning that the criterion of stability in terms of formation energy is not rigorous since some of compositions can decompose against the alternative binary and ternary stable prototypes~\cite{Sokolovskiy-2019}. But these calculations are time-consuming and demanding. Therefore, we restrict our attention to the formation energy calculations only.

\begin{figure}[t] 
\centerline{
\includegraphics[width=10.0cm,clip]{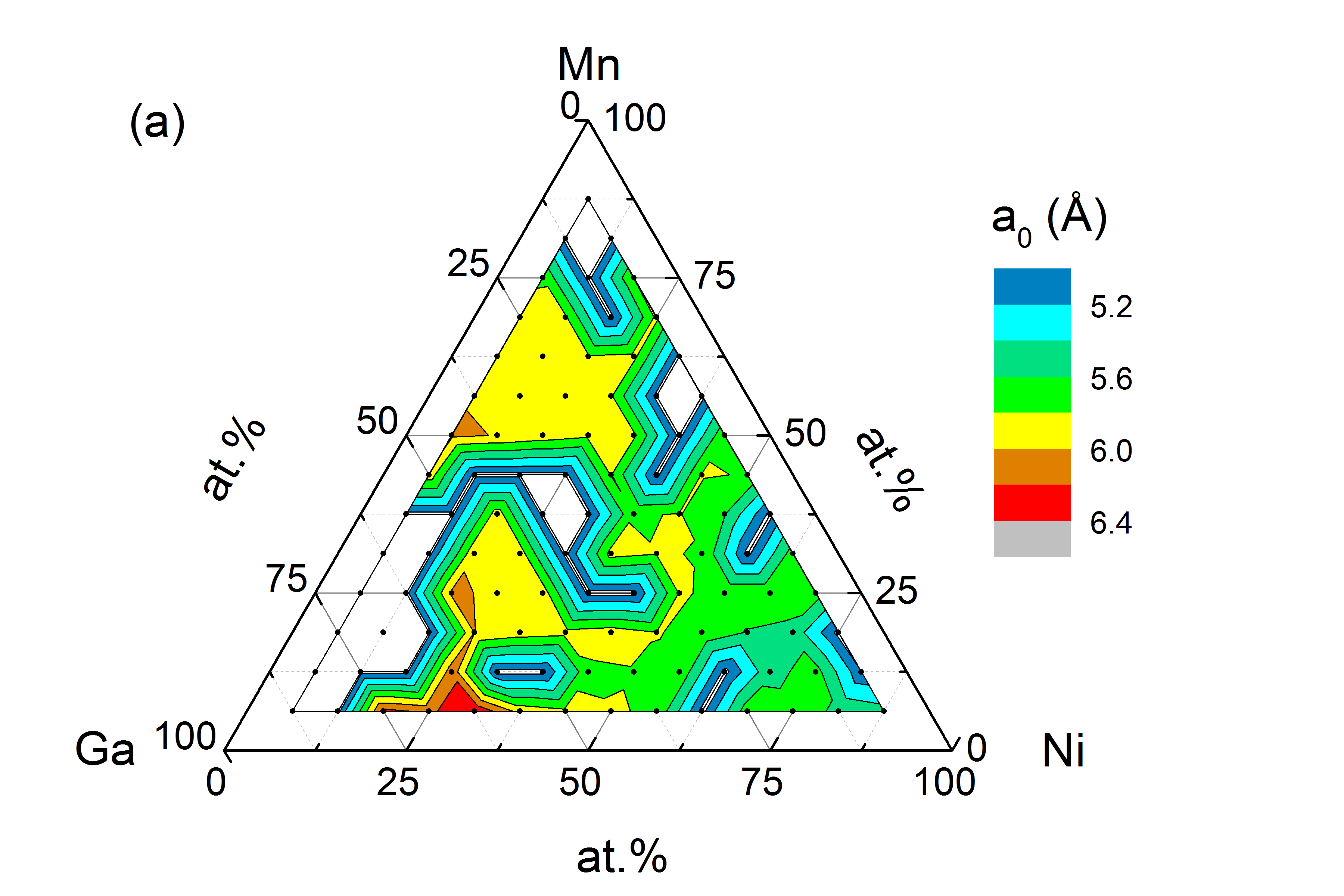}
           }
           \vspace*{0.3cm}
           \centerline{
\includegraphics[width=10.0cm,clip]{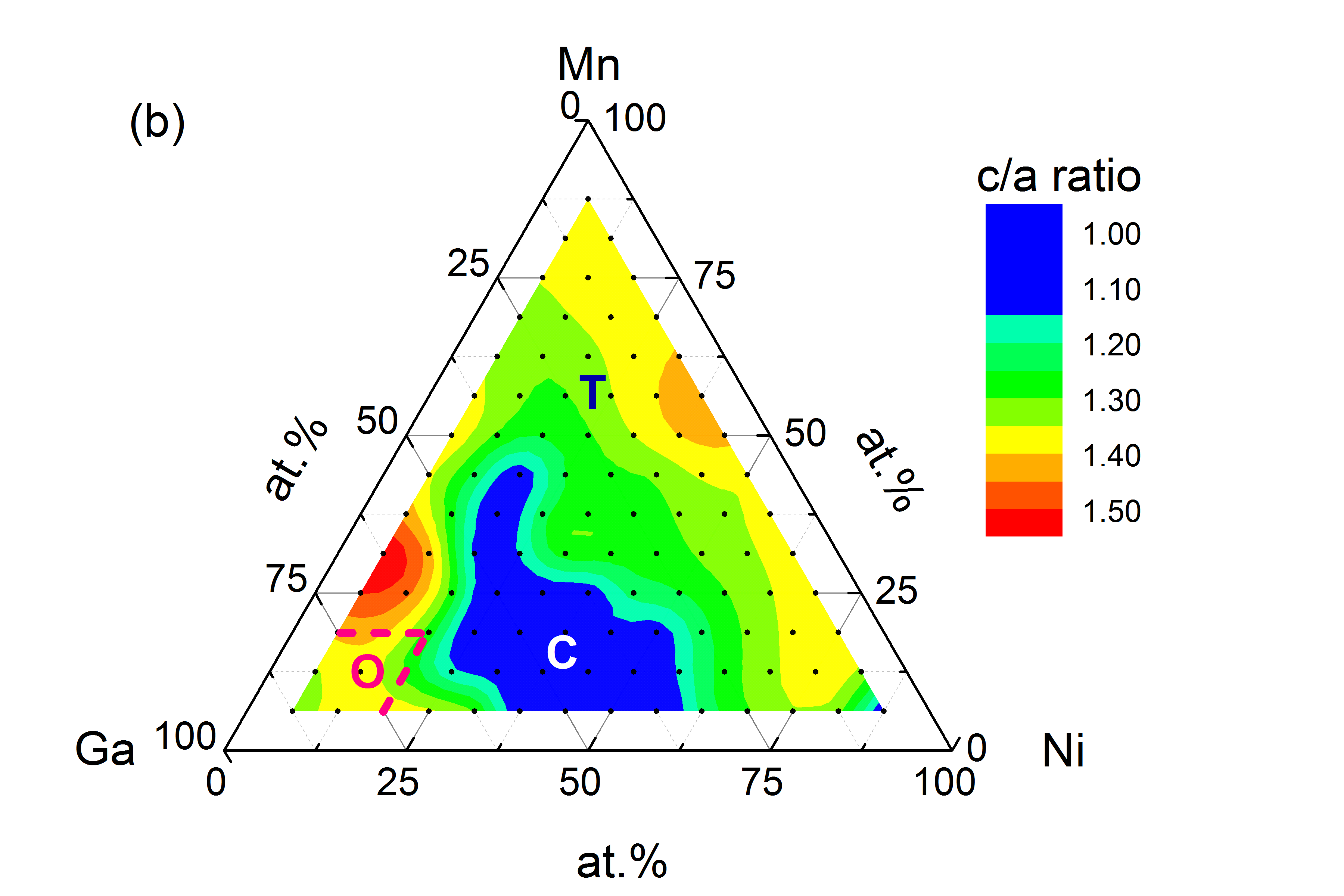}
           }
\caption{\label{fig:epsart} (Color online)
(a) Equilibrium lattice parameter $a_0$ and (b) tetragonal ratio $c/a$ mapped into the ternary diagram for Ni-Mn-Ga compositions in austenite and martensite. Labels C, T and O on the $c/a$ diagram denote the cubic, tetragonal and orthorhombic structure, respectively. For orthorhombic structure, $b/a$ ratio is presented in SM~\cite{SM}.
}
\label{Fig-4}
\end{figure}

 Figure~\ref{Fig-4} illustrates the ternary plots of equilibrium lattice constant $a_0$ of austenite (Fig.~\ref{Fig-4}(a)) and tetragonal ratio $c/a$ of martensite (Fig.~\ref{Fig-4}(b)) in their favorable crystal structures. 
 Notice, for compounds with unstable austenite structure, the lattice constant $a$ is displayed for martensite structure. In addition to that for several Ga- and Ni-rich compositions, martensitic phase is relaxed to the orthorhombic structure. For these compounds the $c/a$ ratio is mapped only.    
  The calculated lattice constants for both austenite and martensite structures are summarized in SM~\cite{SM}. Notice that the lattice constant for the stoichiometric Ni$_2$MnGa, Mn$_2$NiGa, and Ga$_2$MnNi are found to be $\approx$5.81, 5.84, and 5.95~\AA, respectively. These values are close to experimental ones (5.82~\AA~for Ni$_2$MnGa~\cite{Webster-1984}, 5.9~\AA~for Mn$_2$MnGa~\cite{Liu-2005}, and 5.84~\AA~for Ga$_2$MnNi~\cite{Barman-2008}).
   The compositions possessing the largest lattice parameter $a_0$ of austenite structure are located at the left part of ternary diagram with Ga excess content. As Ga content decreases, the lattice constant is found to reduces also, and the smallest values of $a_0$ are observed for Mn- and Ni-rich compositions placed at the top and right corners of diagram, as shown in Fig.~\ref{Fig-4}(a). These findings are connected with the larger atomic radius of Ga in comparison with that for Ni and Mn, while for the latter atomic radii are almost close to each other. Concerning the martensitic phase, it is seen from 
  Fig.~\ref{Fig-4}(b)  that compositions with a high Ga content ($z \gtrsim 85$~at.\%) and a low Ni content ($x \lesssim 7$~at.\%) are stable in the orthorhombic structure with $0.9 < b/a < 1$ and  $1.3 < c/a < 1.4$. Ga-rich compounds with the tetragonal structure have the largest values of $c/a$ ratio in the range of $1.4 < c/a < 1.5$, while Mn- and Ni-rich compounds show $c/a$ ratios in the range of $1.3 < c/a < 1.4$ and  $1.2 < c/a < 1.3$, respectively. Moreover, compositions with Ga ($ 31 < z < 56$ at.\%) and Mn ($y < 35$ at.\%) are stable only in the austenitic cubic phase ($c/a = 1$).
  It should be mentioned that we did not consider the three corners of the ternary diagram represent pure Ni, Mn, and Ga. Nevertheless, the data of tetragonality for the limiting compositions Ni$_{87.5}$Mn$_{6.25}$Ga$_{6.25}$, Mn$_{87.5}$Ni$_{6.25}$Ga$_{6.25}$, and 
  Ga$_{87.5}$Mn$_{6.25}$Ni$_{6.25}$ argues qualitatively with pure elements. Thus, the Ni-rich compound has the cubic structure ($c/a = 1$), the Ga-rich compound show the the orthorhombic structure ($b/a < 1$ and $c/a > 1$), while the Mn-rich compound has the tetragonal structure ($c/a$ = 1.36). Actually, the complex ground state of pure Mn has been discussed in Refs.~\cite{Hobbs-2001,Hobbs-2003,Pulkkinen-2020}.
  
\begin{figure}[t] 
\centerline{
\includegraphics[width=10.0cm,clip]{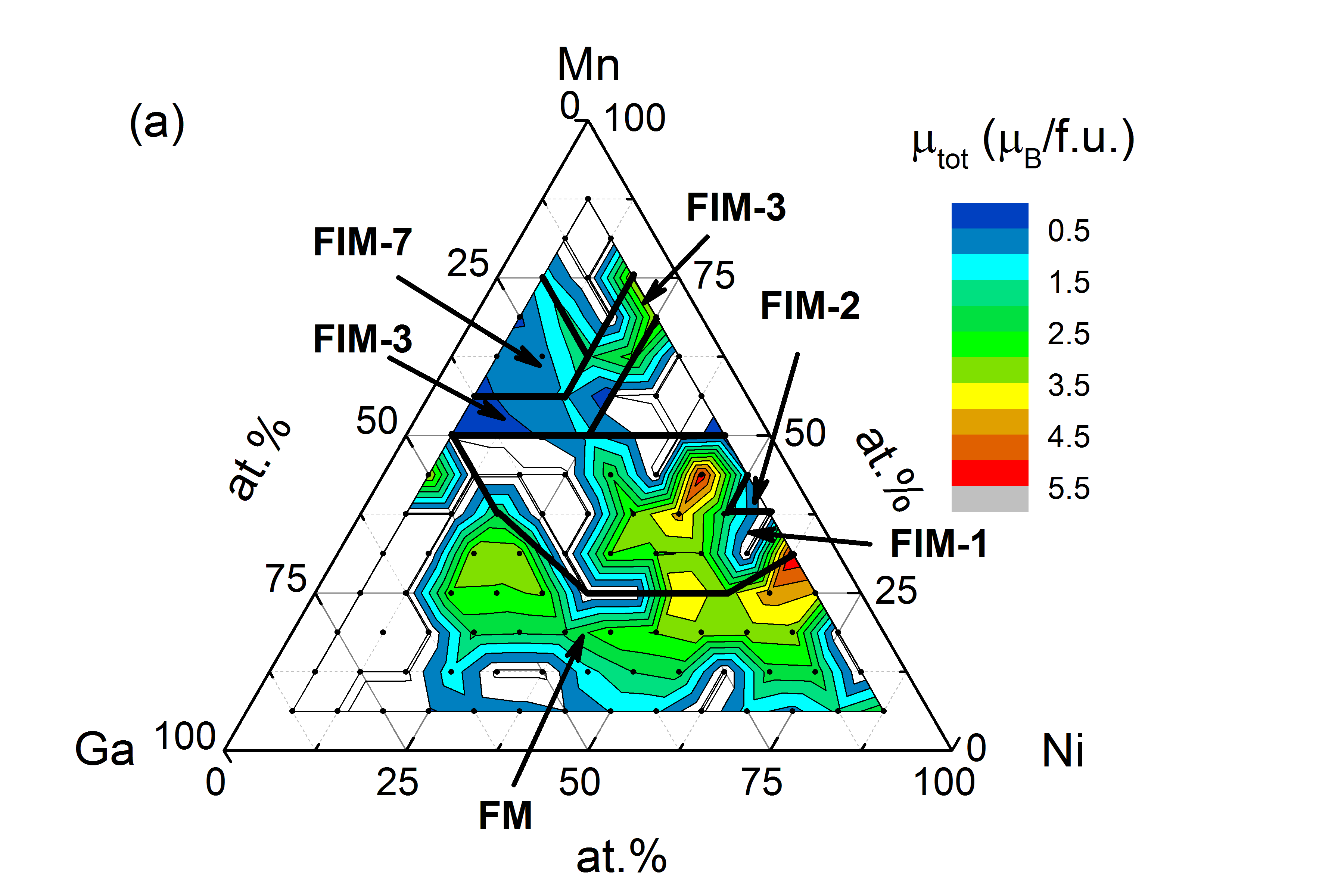}
           }
           \vspace*{0.3cm}
           \centerline{
\includegraphics[width=10.0cm,clip]{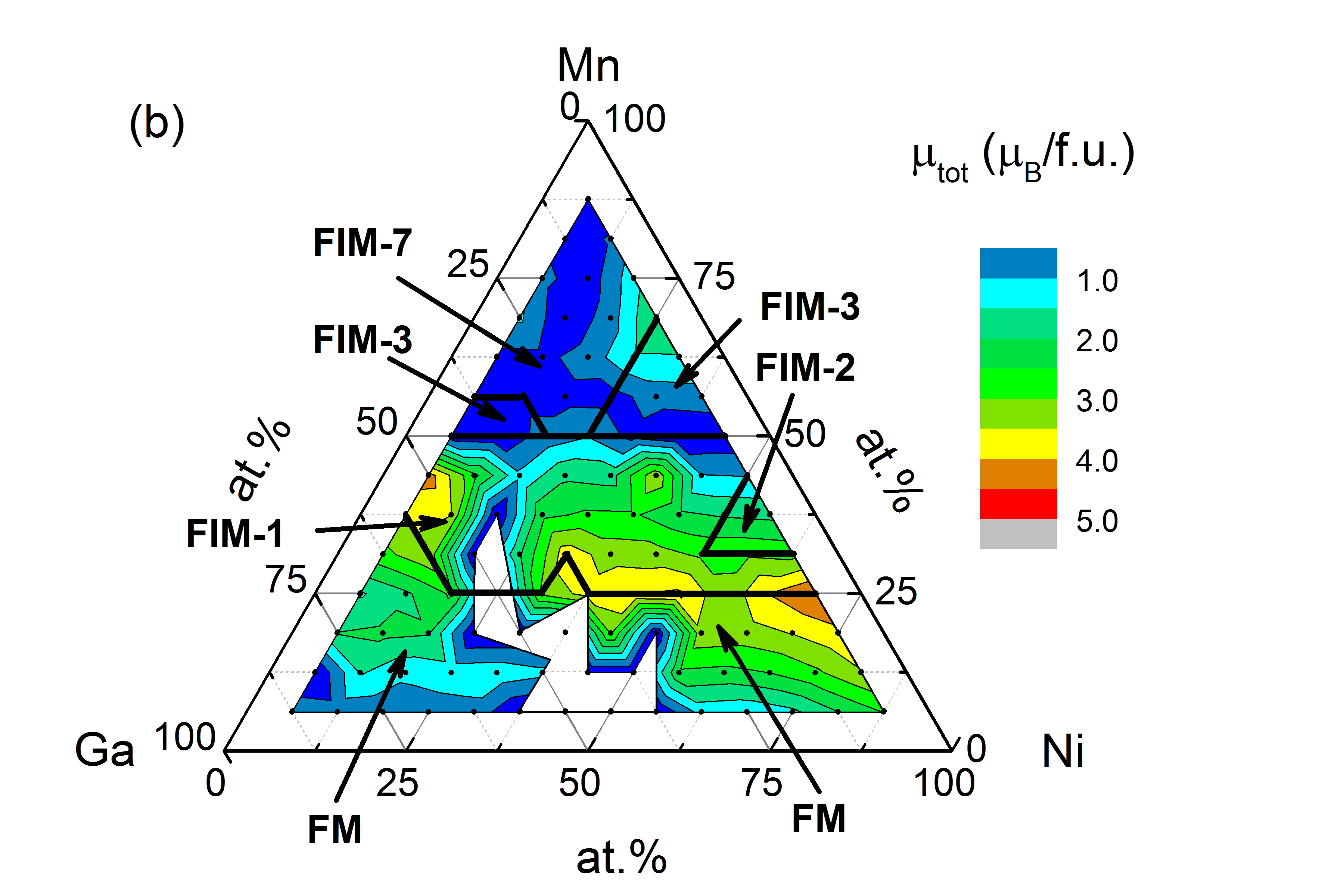}
           }
\caption{\label{fig:epsart} (Color online)
Total magnetic moments and magnetic phase diagram into the ternary diagram for Ni-Mn-Ga compositions in (a) austenite and (b) martensite. White-colored regions indicate the unstable compositions in austenitic and martensitic phases.
}
\label{Fig-5}
\end{figure}

In Fig.~\ref{Fig-5} we demonstrate the mapping of total magnetic moment for stable compositions in the austenite and martensite. Here, we marked additionally areas with favorable magnetic configurations. On the whole, there is a vague similarity between 
these results obtained for austenite and martensite. Namely, compositions with Ga and Mn excess atoms concentrated in the areas I and III possess a small magnetic moment in contrast to the compositions with Ni excess atoms. For compositions from the area I, small values of magnetic moment are concerned with a large amount of non-magnetic Ga atoms that dilutes the ferromagnetically-ordered subsystem. Whereas for compositions from the area III, small magnetic moments are due to an opposite spin orientation between Mn atoms located at different sublattices. 
The largest magnetic moments are found 
to be 5.5$\mu_B$/f.u. and 4.52$\mu_B$/f.u.
for austenite Ni$_{62.5}$Mn$_{31.25}$Ga$_{6.25}$  and  martensite Ni$_{68.75}$Mn$_{25}$Ga$_{6.25}$, which are located in the area II. Besides, the largest difference between magnetic moments in austenite and martensite is found to be $\approx$ 2.5 $\mu_B$/f.u. for compositions from the area II ($50 < x_{\mathrm{Ni}} < 75$~at.\% and $z_{\mathrm{Ga}} < 25$~at.\%) and the area III ($10 < x_{\mathrm{Ni}} < 35$~at.\% and $z_{\mathrm{Ga}} < 25$~at.\%).

\begin{figure}[t] 
\centerline{
\includegraphics[width=10.0cm,clip]{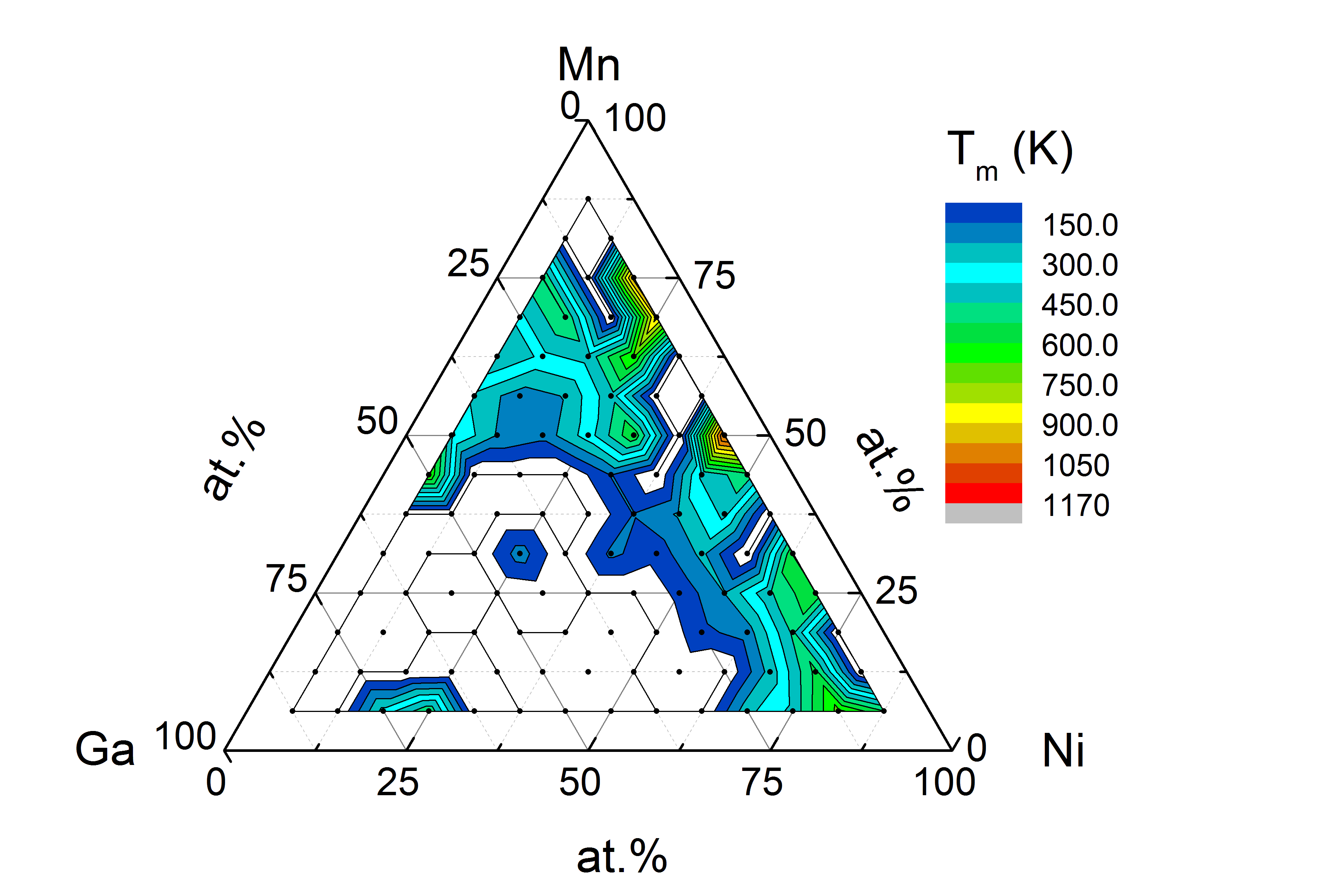}
           }
\caption{\label{fig:epsart} (Color online)
Martensitic transition temperature mapped into the ternary diagram of Ni-Mn-Ga.
White-colored regions indicate the unstable compositions in austenitic phase.}
\label{Fig-6}
\end{figure}

The knowledge of total energy for the favorable austenitic and martensitic phases allows us to estimate the martensitic transition for compositions under study by means of a crude approach $\Delta E \approx k_BT_m$. Here $\Delta E$ is the energy difference between austenite and martensite, $k_B$ is the Boltzmann constant, and $T_m$ is the martensitic transition temperature. The contour map of $T_m$ distribution is depicted in Fig.~\ref{Fig-6}. As evident from the figure, the $T_m$ temperature increases with decreasing Ga content (increasing $e/a$) and thereby that approves the experimental behavior of $T_m(e/a)$.
Conventionally, we can assume that for compositions with Ga content $z > 25$~at.\%, martensitic transition temperature can take place at temperatures less the room temperature.
The smallest values of $T_m$ are found for compositions from the areas I and II, while the largest ones are observed for compositions concentrated along the right side of ternary diagram. 
A correlation between Fig.~\ref{Fig-4}(b) and Fig.~\ref{Fig-6} allows us to conclude that a high martensitic transition temperature is observed for compositions with a large tetragonality ($c/a$ ratio) of the martensitic phase.

\section{Conclusions}
 
 In the present work, the magnetic and structural properties of 105 Ni-Mn-Ga compositions were systematically explored in the framework of first-principles DFT calculations with 16-atom supercell approach. The main results were mapped into the ternary diagrams. The favorable atomic and magnetic configurations for austenitic and martensitic phase were also determined. It was shown that 40 compositions concentrated essentially in the second half of left side and the first half of right side of ternary diagram are found to be unstable in the austenitic phase. Whereas for the martensitic phase, it was found that the compositions with a high Ga content have an orthorhombic structure while the remain compositions possess a tetragonal structure instead of the low-middle part of diagram. For the latter compositions, the martensitic phase was to be unfavorable. Concerning the magnetic phase diagrams, they were observed to be similar with a ferromagnetic and complex ferrimagnetic order for compositions in austenite and martensite. It was shown that the most compositions with Mn content $y_{\mathrm{Mn}} > 25$~at.\% are ordered ferrimagnetically. The present calculated material properties can be useful for design of novel compositions which could have specific properties. However, the prediction problem of a structural stability for such compositions should be expanded to the study of segregation processes also, which can occur in Heusler alloys. But, these calculations are time-consuming and beyond the scope of present research.  
  Such systematic study of ternary Ni-Mn-based Heusler systems would be highly welcome.
  
\section{Supplementary Material}
Ground state properties of stable austenitic and martensitic phases, as well as martensitic transition temperature $T_m$ for Ni-Mn-Ga considered compositions.

\section*{\label{sec:level1}Acknowledgments}
This work was supported by the
Russian Foundation for Basic Research No. 18-32-00507. VS and MZ acknowledge Russian Foundation for Basic Research grant No. 18-08-01434-A.


\end{document}



\preprint{APS/123-QED}

\title{Supplementary material for \\"A ternary map of Ni-Mn-Ga Heusler alloys from \textit{ab initio} calculations"}
\author{Yu. Sokolovskaya}%
 
\affiliation{ 
Faculty of Physics, Chelyabinsk State
University, Chelyabinsk, 454001, Russia}%
\author{O. Miroshkina}
 \affiliation{
Faculty of Physics, Chelyabinsk State
University, Chelyabinsk, 454001, Russia}%
\author{V. Sokolovskiy}%
\affiliation{ 
Faculty of Physics, Chelyabinsk State
University, Chelyabinsk, 454001, Russia}%
\author{M. Zagrebin}
\email{miczag@mail.ru}
 \affiliation{
Faculty of Physics, Chelyabinsk State
University, Chelyabinsk, 454001, Russia}%
 \affiliation{
South Ural State University (national research university), Chelyabinsk, 454080, Russia}

\author{V. Buchelnikov}
 \affiliation{
Faculty of Physics, Chelyabinsk State
University, Chelyabinsk, 454001, Russia}%

\author{A. Zayak}
 \affiliation{
Bowling Green State University, Bowling Green, OH 43403, USA}%

\maketitle

The Wyckoff positions for cubic structures with different atomic arrangement are presented in Table~\ref{tab-1}. 
The magnetic configurations specified by sign of Mn magnetic moments are listed in Table~\ref{tab-2}.

\begin{table}[htb]
    \centering
    \caption{Positions of the Ni, Mn and Ga atoms in the different crystal structures used in our calculations.}
    \label{tab-1}       
    \begin{tabular}{cc|cccc}
    \hline
    \multicolumn{2}{c|}{Wyckoff position} & \#225 & \#{225}$^{\prime}$&\#{225}$^{\prime\prime}$&\#216 \\
    \hline
    4\textit{a} & $\left(0,0,0\right)$    &  Ga  & Ni & Ni & Mn  \\
    4\textit{b}                                       & $\left(\displaystyle\frac{1}{2},\frac{1}{2},\frac{1}{2}\right)$ &  Mn & Mn & Mn & Ni \\[3mm]
    \multirow{ 3}{*}{8\textit{c}$^{*}$} & $\left(\displaystyle\frac{1}{4},\frac{1}{4},\frac{1}{4}\right)$ & Ni   & Ga & Ga/Ni   & Mn  \\ [3mm]

                                                                  & $\left(\displaystyle\frac{3}{4},\frac{3}{4},\frac{3}{4}\right)$ & Ni   & Ga & Ga/Ni   & Ga\\[3mm]
    \hline
    \end{tabular}
    \begin{flushleft}
$^{*}$ For \#216 crystal structure 8\textit{c} Wyckoff position split into two positions: 4\textit{c}(1/4,1/4,1/4) and         4\textit{d} (3/4,3/4,3/4).
\end{flushleft}
\end{table}

\begin{table}[htb]
\centering
\caption{The sign of the magnetic moment of Mn atoms placed at various positions of different crystal structures used in our calculations.}
\label{tab-2}       
\begin{tabular}{l|cccc}
\hline
 & $\left(0,0,0\right)$ &$\left(\displaystyle\frac{1}{4},\frac{1}{4},\frac{1}{4}\right)$&$\left(\displaystyle\frac{1}{2},\frac{1}{2},\frac{1}{2}\right)$& $\left(\displaystyle\frac{3}{4},\frac{3}{4},\frac{3}{4}\right)$
 
 \\[3mm]
 \hline
FM & $>$0 & $>$0 & $>$0 &$>$0 \\
FIM-1 & $>$0 & $<$0 & $>$0 &$<$0 \\
FIM-2 & $<$0 & $>$0 & $>$0 &$>$0 \\
FIM-3 &  $<$0 & $>$0 & $>$0 &$>$0 \\
FIM-4 &  $>$0 & $>$0 & $<$0 &$>$0 \\
FIM-5&  $>$0 & $>$0 & $>$0 &$<$0 \\
FIM-6 & $>$0 & $>$0 & $<$0 &$<$0 \\
FIM-7 & $<$0 & $>$0 & $<$0 &$>$0 \\
FIM-8 &  $<$0 & $>$0 & $>$0 &$<$0 \\
FIM-9 & $<$0 & $>$0 & $<$0 &$<$0\\\hline
\end{tabular}
\end{table}

\vspace{0.5cm}
Ground state properties of stable austenitic and martensitic phases of Ni-Mn-Ga compositions are listed in Tables~\ref{aust_Area_I}-\ref{aust_Area_III} and Tables~\ref{mart_Area_I}-\ref{mart_Area_III}, correspondingly.
Note, empty cells denote the unstable compositions.
The martensitic transition temperatures $T_m$ are summarized in Table~\ref{table_T_m}.

\begin{table}[!h]
    \caption{Lattice parameters $a$, $b$, $c$ and their ratios, the total energy $E_{tot}$, the total magnetic moment $\mu$, the formation energy $E_{\mathrm{form}}$ as well as structure~\# and favorable magnetic state for austenitic compositions of area~I with different concentrations of Ni, Mn, and Ga.}

    \begin{tabular}{ccc|ccc|cc|ccc|c|c}
    \multicolumn{3}{c|}{Concentration [at.\%]} & \multicolumn{3}{c|}{Lattice parameters [\AA]} & \multirow{2}{*}{$b/a$} &  \multirow{2}{*}{$c/a$} &\multirow{2}{*}{$E_{tot}$ [eV/at]}  & \multirow{2}{*}{$\mu$ [$\mu_B$/f.u.]} & \multirow{2}{*}{$E_{\mathrm{form}}$ [eV/f.u.]} &  \multirow{2}{*}{Structure} & \multirow{2}{*}{Mag. State}\\ 
    Ni & Mn & Ga & $a$ & $b$ & $c$ &  &  &  &  &  &  &  \\
     \hline
6.25  & 6.25  & 87.5  &       &       &       &       &       &        &       &        &       &    \\
6.25  & 12.5  & 81.25 &       &       &       &       &       &        &       &        &       &    \\
6.25  & 18.75 & 75    &       &       &       &       &       &        &       &        &       &    \\
6.25  & 25    & 68.75 &       &       &       &       &       &        &       &        &       &    \\
12.5  & 6.25  & 81.25 &       &       &       &       &       &        &       &        &       &    \\
12.5  & 12.5  & 75    &       &       &       &       &       &        &       &        &       &    \\
12.5  & 18.75 & 68.75 &       &       &       &       &       &        &       &        &       &    \\
12.5  & 25    & 62.5  &       &       &       &       &       &        &       &        &       &    \\
18.75 & 6.25  & 75    & 6.263 &       &       & 1.000 & 1.000 & -3.838 & 0.426 & -0.040 & 225$^{\prime}$  & FM \\
18.75 & 12.5  & 68.75 &       &       &       &       &       &        &       &        &       &    \\
18.75 & 18.75 & 62.5  &       &       &       &       &       &        &       &        &       &    \\
18.75 & 25    & 56.25 & 6.047 &       & 6.047 & 1.000 & 1.000 & -5.076 & 2.931 & -0.361 & 225$^{\prime}$  & FM \\
25    & 6.25  & 68.75 & 6.158 &       & 6.156 & 1.000 & 1.000 & -4.025 & 0.434 & -0.066 & 225$^{\prime}$  & FM \\
25    & 12.5  & 62.5  & 6.139 & 5.918 & 6.140 & 0.964 & 1.000 & -4.479 & 1.306 & -0.340 & 225$^{\prime}$  & FM \\
25    & 18.75 & 56.25 & 6.002 & 6.002 & 6.002 & 1.000 & 1.000 & -4.923 & 2.146 & -0.573 & 225$^{\prime}$  & FM \\
25    & 25    & 50    & 5.948 & 5.948 & 5.948 & 1.000 & 1.000 & -5.349 & 2.898 & -0.736 & 225$^{\prime}$  & FM \\
31.25 & 6.25  & 62.5  & 6.388 & 6.388 & 6.234 & 1.000 & 0.976 & -4.361 & 0.781 & -0.691 & 225$^{\prime\prime}$ & FM \\
31.25 & 12.5  & 56.25 &       &       &       &       &       &        &       &        &       &    \\
31.25 & 18.75 & 50    & 5.907 & 5.907 & 5.907 & 1.000 & 1.000 & -5.203 & 2.062 & -0.974 & 225$^{\prime\prime}$ & FM \\
31.25 & 25    & 43.75 & 5.906 & 5.906 & 5.906 & 1.000 & 1.000 & -5.538 & 3.063 & -0.772 & 225$^{\prime\prime}$ & FM \\
37.5  & 6.25  & 56.25 & 5.932 & 5.932 & 5.935 & 1.000 & 1.000 & -4.586 & 0.670 & -0.873 & 225$^{\prime\prime}$ & FM \\
37.5  & 12.5  & 50    &       &       &       &       &       &        &       &        &       &    \\
37.5  & 18.75 & 43.75 & 5.866 & 5.866 & 5.864 & 1.000 & 1.000 & -5.375 & 2.136 & -0.941 & 225$^{\prime\prime}$ & FM \\
37.5  & 25    & 37.5  &       &       &       &       &       &        &       &        &       &    \\
43.75 & 6.25  & 50    & 5.846 & 5.846 & 5.846 & 1.000 & 1.000 & -4.875 & 0.508 & -1.307 & 225$^{\prime\prime}$ & FM \\
43.75 & 12.5  & 43.75 & 5.757 & 5.989 & 5.757 & 1.040 & 1.000 & -5.204 & 1.343 & -1.082 & 225$^{\prime\prime}$ & FM \\
43.75 & 18.75 & 37.5  & 5.907 & 5.835 & 5.837 & 0.988 & 0.988 & -5.576 & 2.873 & -1.029 & 225$^{\prime\prime}$ & FM \\
43.75 & 25    & 31.25 &       &       &       &       &       &        &       &        &       &    \\
50    & 6.25  & 43.75 & 5.821 & 5.821 & 5.820 & 1.000 & 1.000 & -5.073 & 1.028 & -1.380 & 225   & FM \\
50    & 12.5  & 37.5  & 5.783 & 5.872 & 5.783 & 1.015 & 1.000 & -5.446 & 2.020 & -1.330 & 225   & FM \\
50    & 18.75 & 31.25 & 5.806 & 5.806 & 5.805 & 1.000 & 1.000 & -5.819 & 3.023 & -1.281 & 225   & FM \\
50    & 25    & 25    & 5.811 & 5.811 & 5.811 & 1.000 & 1.000 & -6.191 & 4.048 & -1.223 & 225   & FM \\
56.25 & 6.25  & 37.5  & 5.773 & 5.770 & 5.814 & 1.000 & 1.007 & -5.228 & 1.105 & -1.284 & 225   & FM \\
56.25 & 12.5  & 31.25 & 5.773 & 5.773 & 5.791 & 1.000 & 1.003 & -5.600 & 2.216 & -1.229 & 225   & FM \\
56.25 & 18.75 & 25    & 5.769 & 5.769 & 5.769 & 1.000 & 1.000 & -5.971 & 3.289 & -1.169 & 225   & FM \\
56.25 & 25    & 18.75 & 5.775 & 5.774 & 5.775 & 1.000 & 1.000 & -5.939 & 3.211 & 0.501  & 225   & FM \\
62.5  & 6.25  & 31.25 &       &       &       &       &       &        &       &        &       &    \\
62.5  & 12.5  & 25    &       &       &       &       &       &        &       &        &       &    \\
62.5  & 18.75 & 18.75 & 5.582 & 6.027 & 5.582 & 1.080 & 1.000 & -6.081 & 3.232 & -0.887 & 225   & FM \\
62.5  & 25    & 12.5  & 5.775 & 5.641 & 5.773 & 0.977 & 1.000 & -6.411 & 4.323 & -0.666 & 225   & FM \\
68.75 & 6.25  & 25    & 5.710 & 5.710 & 5.709 & 1.000 & 1.000 & -5.533 & 1.170 & -1.062 & 225   & FM \\
68.75 & 12.5  & 18.75 & 5.539 & 6.031 & 5.539 & 1.089 & 1.000 & -5.839 & 2.305 & -0.742 & 225   & FM \\
68.75 & 18.75 & 12.5  & 5.525 & 6.037 & 5.521 & 1.093 & 0.999 & -6.192 & 3.308 & -0.613 & 225   & FM \\
68.75 & 25    & 6.25  & 5.696 & 5.696 & 5.695 & 1.000 & 1.000 & -6.515 & 3.998 & -0.364 & 225   & FM \\
75    & 6.25  & 18.75 & 5.683 & 5.683 & 5.681 & 1.000 & 1.000 & -5.616 & 0.817 & -0.674 & 225   & FM \\
75    & 12.5  & 12.5  & 5.681 & 5.683 & 5.679 & 1.000 & 1.000 & -5.951 & 2.356 & -0.472 & 225   & FM \\
75    & 18.75 & 6.25  &       &       &       &       &       &        &       &        &       &    \\
81.25 & 6.25  & 12.5  & 5.569 & 5.698 & 5.699 & 1.023 & 1.023 & -5.713 & 1.604 & -0.344 & 225   & FM \\
81.25 & 12.5  & 6.25  &       &       &       &       &       &        &       &        &       &    \\
87.5  & 6.25  & 6.25  & 5.479 & 5.909 & 5.483 & 1.078 & 1.001 & -5.815 & 2.386 & -0.032 & 225   & FM 
\end{tabular}
    \label{aust_Area_I}
\end{table}

\begin{table}[!h]
    \caption{Lattice parameters $a$, $b$, $c$ and their ratios, the total energy $E_{tot}$, the total magnetic moment $\mu$, the formation energy $E_{\mathrm{form}}$ as well as structure~\# and favorable magnetic state for austenitic compositions of area~II with different concentrations of Ni, Mn, and Ga.}
    \begin{tabular}{ccc|ccc|cc|ccc|c|c}
    \multicolumn{3}{c|}{Concentration [at.\%]} & \multicolumn{3}{c|}{Lattice parameters [\AA]} & \multirow{2}{*}{$b/a$} &  \multirow{2}{*}{$c/a$} &\multirow{2}{*}{$E_{tot}$ [eV/at]}  & \multirow{2}{*}{$\mu$ [$\mu_B$/f.u.]} & \multirow{2}{*}{$E_{\mathrm{form}}$ [eV/f.u.]} &  \multirow{2}{*}{Structure} & \multirow{2}{*}{Mag. State}\\ 
    Ni & Mn & Ga & $a$ & $b$ & $c$ &  &  &  &  &  &  & \\
    \hline
62.5 & 31.25 & 6.25 & 5.703 & 5.818 & 5.703 & 1.020 & 1.000 & -6.739 & 5.508 & -0.436 & 225 & FM \\
56.25 & 31.25 & 12.5 & & & & & & & & & 225 & \\
56.25 & 37.5 & 6.25 &  &  &  &  &  &  &  &  &  &  \\
50 & 31.25 & 18.75 & 5.794 & 5.794 & 5.790 & 1.000 & 0.999 & -6.518 & 3.057 & -0.990 & 225 & FIM-2 \\
50 & 37.5 & 12.5 & 5.607 & 6.130 & 5.607 & 1.093 & 1.000 & -6.848 & 2.012 & -0.766 & 225 & FIM-2 \\
50 & 43.75 & 6.25 & 5.788 & 5.788 & 5.786 & 1.000 & 1.000 & -7.172 & 0.924 & -0.518 & 225 & FIM-2 \\
43.75 & 31.25 & 25 & 5.808 & 5.808 & 5.807 & 1.000 & 1.000 & -6.367 & 2.957 & -1.103 & 225 & FIM-1 \\
43.75 & 37.5 & 18.75 & 5.811 & 5.811 & 5.810 & 1.000 & 1.000 & -6.692 & 4.177 & -0.863 & 225 & FIM-1 \\
43.75 & 43.75 & 12.5 & 5.819 & 5.819 & 5.821 & 1.000 & 1.000 & -7.021 & 5.341 & -0.634 & 225 & FIM-1 \\
37.5 & 31.25 & 31.25 & 5.875 & 5.860 & 5.858 & 0.998 & 0.997 & -6.138 & 2.999 & -0.907 & 225 & FIM-1 \\
37.5 & 37.5 & 25 & 5.780 & 5.793 & 5.793 & 1.002 & 1.002 & -6.549 & 3.111 & -1.007 & 225 & FIM-1 \\
37.5 & 43.75 & 18.75 & & & & & & & & & & \\
31.25 & 31.25 & 37.5 & & & & & & & & & & \\
31.25 & 37.5 & 31.25 & &  &  & &  &  &  &  &  &  \\
31.25 & 43.75 & 25 & 5.815 & 5.815 & 5.755 & 1.000 & 0.990 & -6.737 & 2.400 & -0.939 & 225 & FIM-1 \\
25 & 31.25 & 43.75 & 5.904 & 5.904 & 5.902 & 1.000 & 1.000 & -5.717 & 3.335 & -0.664 & 225$^{\prime}$ & FM \\
25 & 37.5 & 37.5 & & & & & & & & & & \\
25 & 43.75 & 31.25 & & & & & & & & & & \\
18.75 & 31.25 & 50 & 5.961 & 5.961 & 5.959 & 1.000 & 1.000 & -5.543 & 3.430 & -0.688 & 225$^{\prime}$ & FM \\
18.75 & 37.5 & 43.75 & 5.913 & 5.913 & 5.911 & 1.000 & 1.000 & -5.938 & 2.019 & -0.724 & 225$^{\prime}$ & FIM-1 \\
18.75 & 43.75 & 37.5 & & & & & & & & & & \\
12.5 & 31.25 & 56.25 & & & & & & & & & & \\
12.5 & 37.5 & 50 & & & & & & & & & & \\
12.5 & 43.75 & 43.75 & & & & & & & & & & \\
6.25 & 31.25 & 62.5 & & & & & & & & & & \\
6.25 & 37.5 & 56.25 &   &   &   &   &   &   &  &   &   &  \\
6.25 & 43.75 & 50 & 5.921 & 5.921 & 5.919 & 1.000 & 1.000 & -5.903 & 3.511 & -0.480 & 225$^{\prime}$ & FM 
\end{tabular}
    \label{aust_Area_II}
\end{table}

\begin{table}[]
    \caption{Lattice parameters $a$, $b$, $c$ and their ratios, the total energy $E_{tot}$, the total magnetic moment $\mu$, the formation energy $E_{\mathrm{form}}$ as well as structure~\# and favorable magnetic state for austenitic compositions of area~III with different concentrations of Ni, Mn, and Ga.}
    \begin{tabular}{ccc|ccc|cc|ccc|c|c}
    \multicolumn{3}{c|}{Concentration [at.\%]} & \multicolumn{3}{c|}{Lattice parameters [\AA]} & \multirow{2}{*}{$b/a$} &  \multirow{2}{*}{$c/a$} &\multirow{2}{*}{$E_{tot}$ [eV/at]}  & \multirow{2}{*}{$\mu$ [$\mu_B$/f.u.]} & \multirow{2}{*}{$E_{\mathrm{form}}$ [eV/f.u.]} &  \multirow{2}{*}{Structure} & \multirow{2}{*}{Mag. State}\\ 
    Ni & Mn & Ga & $a$ & $b$ & $c$ &  &  &  &  &  &  & \\
    \hline
    43.75 & 50 & 6.25 & 5.761 & 5.761 & 5.755 & 1.000 & 0.999 & -7.347 & 0.193 & 0.193 & 216 & FIM-3 \\
37.5 & 50 & 12.5 & & & & & & & & & & \\
37.5 & 56.25 & 6.25 & & & & & & & & & & \\
31.25 & 50 & 18.75 & 5.844 & 5.844 & 5.840 & 1.000 & 0.999 & -7.049 & 0.521 & 0.521 & 216 & FIM-3 \\
31.25 & 56.25 & 12.5 & & & & & & & & & & \\
31.25 & 62.5 & 6.25 & & & & & & & & & & \\
25 & 50 & 25 & 5.840 & 5.840 & 5.840 & 1.000 & 1.000 & -6.913 & 1.132 & 1.132 & 216 & FIM-3 \\
25 & 56.25 & 18.75 & 5.827 & 5.827 & 5.827 & 1.000 & 1.000 & -7.236 & 0.105 & 0.105 & 216 & FIM-3 \\
25 & 62.5 & 12.5 & 5.859 & 5.836 & 5.836 & 0.996 & 0.996 & -7.576 & 2.722 & 2.722 & 216 & FIM-3 \\
25 & 68.75 & 6.25 & 5.850 & 5.850 & 5.850 & 1.000 & 1.000 & -7.910 & 3.503 & 3.503 & 216 & FIM-3 \\
18.75 & 50 & 31.25 & 5.917 & 5.917 & 5.917 & 1.000 & 1.000 & -6.692 & 0.557 & 0.557 & 216 & FIM-3 \\
18.75 & 56.25 & 25 & 5.825 & 5.825 & 5.825 & 1.000 & 1.000 & -7.101 & 1.055 & 1.055 & 216 & FIM-7 \\
18.75 & 62.5 & 18.75 & 5.820 & 5.820 & 5.820 & 1.000 & 1.000 & -7.433 & 1.925 & 1.925 & 216 & FIM-7 \\
18.75 & 68.75 & 12.5 & & & & & & & & & & \\
18.75 & 75 & 6.25 & 5.804 & 5.804 & 5.804 & 1.000 & 1.000 & -8.095 & 3.561 & 3.561 & 216 & FIM-3 \\
12.5 & 50 & 37.5 & 5.990 & 5.990 & 5.952 & 1.000 & 0.994 & -6.466 & 0.143 & 0.143 & 216 & FIM-3 \\
12.5 & 56.25 & 31.25 & 5.889 & 5.903 & 5.903 & 1.002 & 1.002 & -6.878 & 0.802 & 0.802 & 216 & FIM-3 \\
12.5 & 62.5 & 25 & 5.830 & 5.830 & 5.796 & 1.000 & 0.994 & -7.284 & 0.786 & 0.786 & 216 & FIM-7 \\
12.5 & 68.75 & 18.75 & 5.828 & 5.828 & 5.759 & 1.000 & 0.988 & -7.618 & 1.720 & 1.720 & 216 & FIM-7 \\
12.5 & 75 & 12.5 & & & & & & & & & & \\
12.5 & 81.25 & 6.25 & & & & & & & & & & \\
6.25 & 50 & 43.75 & 6.042 & 6.042 & 6.040 & 1.000 & 1.000 & -6.224 & 0.111 & 0.111 & 216 & FIM-3 \\
6.25 & 56.25 & 37.5 & 5.977 & 5.977 & 5.946 & 1.000 & 0.995 & -6.649 & 0.253 & 0.253 & 216 & FIM-3 \\
6.25 & 62.5 & 31.25 & 5.915 & 5.915 & 5.829 & 1.000 & 0.985 & -7.054 & 0.720 & 0.720 & 216 & FIM-7 \\
6.25 & 68.75 & 25 & 5.821 & 5.821 & 5.821 & 1.000 & 1.000 & -7.469 & 0.419 & 0.419 & 216 & FIM-7 \\
6.25 & 75 & 18.75 & 5.792 & 5.792 & 5.792 & 1.000 & 1.000 & -7.806 & 1.343 & 1.343 & 216 & FIM-7 \\
6.25 & 81.25 & 12.5 & & & & & & & & & & \\
6.25 & 87.5 & 6.25 & & & & & & & & & & 
\end{tabular}
    \label{aust_Area_III}
\end{table}


\begin{table}[]
    \caption{Lattice parameters $a$, $b$, $c$ and their ratios, the total energy $E_{tot}$, the total magnetic moment $\mu$, the formation energy $E_{\mathrm{form}}$ as well as the favorable magnetic state for martensitic compositions of area~I with different concentrations of Ni, Mn, and Ga.}
    \label{mart_Area_I}
    \begin{tabular}{ccc|ccc|cc|ccc|c}
    \multicolumn{3}{c|}{Concentration [at.\%]} & \multicolumn{3}{c|}{Lattice parameters [\AA]} & \multirow{2}{*}{$b/a$} &  \multirow{2}{*}{$c/a$} &\multirow{2}{*}{$E_{tot}$ [eV/at]}  & \multirow{2}{*}{$\mu$ [$\mu_B$/f.u.]} & \multirow{2}{*}{$E_{\mathrm{form}}$ [eV/f.u.]} & \multirow{2}{*}{Mag. State}\\ 
    Ni & Mn & Ga & $a$ & $b$ & $c$ &  &  &  &   &  & \\
    \hline
6.25  & 6.25  & 87.5  & 5.960 & 5.813 & 7.871 & 0.975 & 1.321 & -3.459 & 0.800  & 0.038  & FM \\
6.25  & 12.5  & 81.25 & 5.993 & 5.521 & 7.865 & 0.921 & 1.312 & -3.859 & -0.113 & -0.019 & FM \\
6.25  & 18.75 & 75    & 5.756 & 5.306 & 8.034 & 0.922 & 1.396 & -4.240 & 2.014  & -0.002 & FM \\
6.25  & 25    & 68.75 & 5.312 &       & 7.952 & 1.000 & 1.497 & -4.687 & 1.438  & -0.243 & FM \\
12.5  & 6.25  & 81.25 & 5.737 &       & 7.893 & 1.000 & 1.376 & -3.683 & 0.704  & -0.140 & FM \\
12.5  & 12.5  & 75    & 5.527 & 5.734 & 7.866 & 1.037 & 1.423 & -4.060 & 1.550  & -0.104 & FM \\
12.5  & 18.75 & 68.75 & 5.486 &       & 7.813 & 1.000 & 1.424 & -4.453 & 2.121  & -0.134 & FM \\
12.5  & 25    & 62.5  & 5.334 &       & 7.853 & 1.000 & 1.472 & -4.898 & 1.913  & -0.369 & FM \\
18.75 & 6.25  & 75    & 5.615 &       & 7.718 & 1.000 & 1.374 & -3.867 & 0.610  & -0.154 & FM \\
18.75 & 12.5  & 68.75 & 5.909 & 5.339 & 7.330 & 0.904 & 1.240 & -4.269 & 1.134  & -0.219 & FM \\
18.75 & 18.75 & 62.5  & 5.554 &       & 7.296 &       & 1.314 & -4.668 & 2.045  & -0.274 & FM \\
18.75 & 25    & 56.25 & 5.412 &       & 7.402 &       & 1.368 & -5.078 & 2.572  & -0.371 & FM \\
25    & 6.25  & 68.75 & 5.487 &       & 7.661 & 1.000 & 1.396 & -4.069 & 0.722  & -0.244 & FM \\
25    & 12.5  & 62.5  & 5.974 & 5.998 & 6.231 & 1.004 & 1.043 & -4.480 & 1.292  & -0.343 & FM \\
25    & 18.75 & 56.25 &       &       &       &       &       &        &        &        &    \\
25    & 25    & 50    & run   &       &       &       &       &        &        &        &    \\
31.25 & 6.25  & 62.5  & 5.475 & 5.475 & 7.752 & 1.000 & 1.416 & -4.362 & 0.810  & -0.697 & FM \\
31.25 & 12.5  & 56.25 & 5.619 & 5.619 & 6.687 & 1.000 & 1.190 & -4.765 & 1.349  & -0.765 & FM \\
31.25 & 18.75 & 50    &       &       &       &       &       &        &        &        &    \\
31.25 & 25    & 43.75 & 5.628 & 5.628 & 6.506 & 1.000 & 1.156 & -5.541 & 3.103  & -0.781 & FM \\
37.5  & 6.25  & 56.25 &       &       &       &       &       &        &        &        &    \\
37.5  & 12.5  & 50    & 5.742 & 5.742 & 6.110 & 1.000 & 1.064 & -5.046 & 1.139  & -1.170 & FM \\
37.5  & 18.75 & 43.75 &       &       &       &       &       &        &        &        &    \\
37.5  & 25    & 37.5  & 6.098 & 5.813 & 5.811 & 0.953 & 0.953 & -5.732 & 3.704  & -0.828 & FM \\
43.75 & 6.25  & 50    &       &       &       &       &       &        &        &        &    \\
43.75 & 12.5  & 43.75 &       &       &       &       &       &        &        &        &    \\
43.75 & 18.75 & 37.5  & 5.397 & 5.397 & 6.897 & 1.000 & 1.278 & -5.576 & 3.036  & -1.029 & FM \\
43.75 & 25    & 31.25 & 5.373 & 5.373 & 6.900 & 1.000 & 1.284 & -5.972 & 3.980  & -1.067 &    \\
50    & 6.25  & 43.75 &       &       &       &       &       &        &        &        &    \\
50    & 12.5  & 37.5  &       &       &       &       &       &        &        &        &    \\
50    & 18.75 & 31.25 &       &       &       &       &       &        &        &        &    \\
50    & 25    & 25    & 5.380 & 5.380 & 6.759 & 1.000 & 1.256 & -6.199 & 4.088  & -1.256 & FM \\
56.25 & 6.25  & 37.5  &       &       &       &       &       &        &        &        &    \\
56.25 & 12.5  & 31.25 & 5.424 & 5.424 & 6.552 & 1.000 & 1.208 & -5.602 & 2.421  & -1.237 & FM \\
56.25 & 18.75 & 25    & 5.354 & 5.354 & 6.689 & 1.000 & 1.249 & -5.982 & 3.354  & -1.213 & FM \\
56.25 & 25    & 18.75 & 5.301 & 5.279 & 6.848 & 0.996 & 1.292 & -5.959 & 3.094  & 0.423  & FM \\
62.5  & 6.25  & 31.25 & 5.324 & 5.324 & 6.681 & 1.000 & 1.255 & -5.402 & 1.199  & -1.260 & FM \\
62.5  & 12.5  & 25    & 5.289 & 5.289 & 6.727 & 1.000 & 1.272 & -5.782 & 2.349  & -1.236 & FM \\
62.5  & 18.75 & 18.75 & 5.230 & 5.274 & 6.814 & 1.008 & 1.303 & -6.097 & 3.289  & -0.951 & FM \\
62.5  & 25    & 12.5  & 5.172 & 5.189 & 6.952 & 1.003 & 1.344 & -6.449 & 4.023  & -0.818 & FM \\
68.75 & 6.25  & 25    & 5.238 & 5.238 & 6.758 & 1.000 & 1.290 & -5.556 & 1.113  & -1.157 & FM \\
68.75 & 12.5  & 18.75 & 5.188 & 5.226 & 6.851 & 1.007 & 1.321 & -5.861 & 2.508  & -0.831 & FM \\
68.75 & 18.75 & 12.5  & 5.128 & 5.167 & 6.969 & 1.008 & 1.359 & -6.232 & 3.521  & -0.774 & FM \\
68.75 & 25    & 6.25  & 5.100 & 5.100 & 7.075 & 1.000 & 1.387 & -6.567 & 4.529  & -0.570 & FM \\
75    & 6.25  & 18.75 & 5.141 & 5.141 & 6.916 & 1.000 & 1.345 & -5.650 & 1.388  & -0.811 & FM \\
75    & 12.5  & 12.5  & 5.108 & 5.118 & 6.981 & 1.002 & 1.367 & -5.998 & 2.900  & -0.661 & FM \\
75    & 18.75 & 6.25  & 5.050 & 5.077 & 7.086 & 1.005 & 1.403 & -6.357 & 3.857  & -0.554 & FM \\
81.25 & 6.25  & 12.5  & 5.044 & 5.064 & 7.051 & 1.004 & 1.398 & -5.773 & 1.991  & -0.586 & FM \\
81.25 & 12.5  & 6.25  & 5.055 & 5.055 & 7.051 & 1.000 & 1.395 & -6.110 & 3.446  & -0.387 & FM \\
87.5  & 6.25  & 6.25  & 5.201 & 6.529 & 5.154 & 1.255 & 0.991 & -5.860 & 2.518  & -0.214 & FM
\end{tabular}
\end{table}

\begin{table}[]
    \caption{Lattice parameters $a$, $b$, $c$ and their ratios, the total energy $E_{tot}$, the total magnetic moment $\mu$, the formation energy $E_{\mathrm{form}}$ as well as the favorable magnetic state for martensitic compositions of area~II with different concentrations of Ni, Mn, and Ga.}
    \label{mart_Area_II}
    \begin{tabular}{ccc|ccc|cc|ccc|c}
    \multicolumn{3}{c|}{Concentration [at.\%]} & \multicolumn{3}{c|}{Lattice parameters [\AA]} & \multirow{2}{*}{$b/a$} &  \multirow{2}{*}{$c/a$} &\multirow{2}{*}{$E_{tot}$ [eV/at]}  & \multirow{2}{*}{$\mu$ [$\mu_B$/f.u.]} & \multirow{2}{*}{$E_{\mathrm{form}}$ [eV/f.u.]} & \multirow{2}{*}{Mag. State}\\ 
    Ni & Mn & Ga & $a$ & $b$ & $c$ &  &  &  &   &  & \\
    \hline
    62.5 & 31.25 & 6.25 & 5.130 & 5.139 & 7.023 & 1.002 & 1.369 & -6.791 & 2.837 & -0.644 & FIM-2 \\
56.25 & 31.25 & 12.5 & 5.202 & 5.202 & 6.985 & 1.000 & 1.343 & -6.665 & 2.918 & -0.859 & FIM-2 \\
56.25 & 37.5 & 6.25 & 5.155 & 5.156 & 7.053 & 1.000 & 1.368 & -7.010 & 1.873 & -0.694 & FIM-2 \\
50 & 31.25 & 18.75 & 5.276 & 5.276 & 6.916 & 1.000 & 1.311 & -6.539 & 2.903 & -1.074 & FIM-2 \\
50 & 37.5 & 12.5 & 5.211 & 5.220 & 7.029 & 1.002 & 1.349 & -6.882 & 1.943 & -0.903 & FIM-2 \\
50 & 43.75 & 6.25 & 5.209 & 5.209 & 6.904 & 1.000 & 1.325 & -7.215 & 0.973 & -0.691 & FIM-2 \\
43.75 & 31.25 & 25 & 5.363 & 5.363 & 6.767 & 1.000 & 1.262 & -6.376 & 3.220 & -1.139 & FIM-1 \\
43.75 & 37.5 & 18.75 & 5.293 & 5.293 & 6.914 & 1.000 & 1.306 & -6.713 & 2.104 & -0.944 & FIM-1 \\
43.75 & 43.75 & 12.5 & 5.193 & 5.193 & 7.105 & 1.000 & 1.368 & -7.054 & 1.017 & -0.768 & FIM-1 \\
37.5 & 31.25 & 31.25 & 5.388 & 5.368 & 6.940 & 0.996 & 1.288 & -6.152 & 3.288 & -0.964 & FIM-1 \\
37.5 & 37.5 & 25 & 5.353 & 5.374 & 6.781 & 1.004 & 1.267 & -6.562 & 2.564 & -1.061 & FIM-1 \\
37.5 & 43.75 & 18.75 & 5.352 & 5.350 & 6.802 & 1.000 & 1.271 & -6.892 & 3.370 & -0.837 & FM \\
31.25 & 31.25 & 37.5 & 5.281 & 5.281 & 7.287 & 1.000 & 1.380 & -5.929 & 3.783 & -0.790 & FM \\
31.25 & 37.5 & 31.25 & 5.405 & 5.368 & 6.927 & 0.993 & 1.282 & -6.341 & 2.587 & -0.896 & FIM-1 \\
31.25 & 43.75 & 25 & 5.347 & 5.347 & 6.860 & 1.000 & 1.283 & -6.751 & 1.805 & -0.993 & FIM-1 \\
25 & 31.25 & 43.75 & 5.573 & 5.573 & 6.650 & 1.000 & 1.193 & -5.733 & 1.924 & -0.729 & FIM-1 \\
25 & 37.5 & 37.5 & 5.407 & 5.407 & 7.130 & 1.000 & 1.319 & -6.132 & 2.387 & -0.780 & FIM-1 \\
25 & 43.75 & 31.25 & 5.369 & 5.369 & 6.993 & 1.000 & 1.303 & -6.536 & 1.900 & -0.855 & FIM-1 \\
18.75 & 31.25 & 50 & & & & & & 0.000 & 0.000 & 21.485 & \\
18.75 & 37.5 & 43.75 & & & & & & 0.000 & 0.000 & 23.028 & \\
18.75 & 43.75 & 37.5 & 6.107 & 5.783 & 5.783 & 0.947 & 0.947 & -6.321 & 1.108 & -0.713 & FM \\
12.5 & 31.25 & 56.25 & 5.211 & 5.283 & 8.017 & 1.014 & 1.538 & -5.323 & 3.409 & -0.528 & FM \\
12.5 & 37.5 & 50 & 5.442 & 5.442 & 7.107 & 1.000 & 1.306 & -5.764 & 3.501 & -0.747 & FIM-2 \\
12.5 & 43.75 & 43.75 & 5.433 & 5.433 & 7.028 & 1.000 & 1.293 & -6.147 & 2.639 & -0.736 & FIM-1 \\
6.25 & 31.25 & 62.5 & 5.243 & 5.243 & 8.108 & 1.000 & 1.546 & -5.116 & 2.848 & -0.416 & FM \\
6.25 & 37.5 & 56.25 & 5.383 & 5.397 & 7.534 & 1.003 & 1.400 & -5.515 & 3.582 & -0.471 & FM \\
6.25 & 43.75 & 50 & 5.423 & 5.423 & 7.227 & 1.000 & 1.333 & -5.962 & 4.228 & -0.715 & FM 
\end{tabular}
\end{table}

\begin{table}[]
    \caption{Lattice parameters $a$, $b$, $c$ and their ratios, the total energy $E_{tot}$, the total magnetic moment $\mu$, the formation energy $E_{\mathrm{form}}$ as well as the favorable magnetic state for martensitic compositions of area~III with different concentrations of Ni, Mn, and Ga.}
    \label{mart_Area_III}
    \begin{tabular}{ccc|ccc|cc|ccc|c}
    \multicolumn{3}{c|}{Concentration [at.\%]} & \multicolumn{3}{c|}{Lattice parameters [\AA]} & \multirow{2}{*}{$b/a$} &  \multirow{2}{*}{$c/a$} &\multirow{2}{*}{$E_{tot}$ [eV/at]}  & \multirow{2}{*}{$\mu$ [$\mu_B$/f.u.]} & \multirow{2}{*}{$E_{\mathrm{form}}$ [eV/f.u.]} & \multirow{2}{*}{Mag. State}\\ 
    Ni & Mn & Ga & $a$ & $b$ & $c$ &  &  &  &   &  & \\
    \hline
    43.75 & 50 & 6.25 & 5.105 & 5.105 & 7.336 & 1.000 & 1.437 & -7.447 & 0.113 & -0.797 & FIM-3 \\
37.5 & 50 & 12.5 & 5.141 & 5.155 & 7.330 & 1.003 & 1.426 & -7.276 & 0.027 & -0.832 & FIM-3 \\
37.5 & 56.25 & 6.25 & 5.131 & 5.117 & 7.232 & 0.997 & 1.410 & -7.624 & 0.769 & -0.680 & FIM-3 \\
31.25 & 50 & 18.75 & 5.251 & 5.251 & 7.154 & 1.000 & 1.362 & -7.100 & 0.326 & -0.848 & FIM-3 \\
31.25 & 56.25 & 12.5 & 5.162 & 5.169 & 7.247 & 1.001 & 1.404 & -7.458 & 0.715 & -0.736 & FIM-3 \\
31.25 & 62.5 & 6.25 & 5.145 & 5.142 & 7.191 & 0.999 & 1.398 & -7.806 & 1.455 & -0.586 & FIM-3 \\
25 & 50 & 25 & 5.362 & 5.362 & 6.868 & 1.000 & 1.281 & -6.942 & 0.977 & -0.936 & FIM-3 \\
25 & 56.25 & 18.75 & 5.267 & 5.267 & 7.123 & 1.000 & 1.352 & -7.273 & 0.305 & -0.715 & FIM-3 \\
25 & 62.5 & 12.5 & 5.185 & 5.185 & 7.185 & 1.000 & 1.386 & -7.636 & 1.431 & -0.622 & FIM-3 \\
25 & 68.75 & 6.25 & 5.155 & 5.155 & 7.156 & 1.000 & 1.388 & -7.990 & 2.044 & -0.497 & FIM-3 \\
18.75 & 50 & 31.25 & 5.416 & 5.416 & 6.959 & 1.000 & 1.285 & -6.707 & 0.662 & -0.713 & FIM-3 \\
18.75 & 56.25 & 25 & 5.378 & 5.378 & 6.894 & 1.000 & 1.282 & -7.121 & 0.329 & -0.827 & FIM-7 \\
18.75 & 62.5 & 18.75 & 5.282 & 5.282 & 7.009 & 1.000 & 1.327 & -7.468 & 0.724 & -0.669 & FIM-7 \\
18.75 & 68.75 & 12.5 & 5.182 & 5.180 & 7.161 & 1.000 & 1.382 & -7.823 & 0.770 & -0.547 & FIM-7 \\
18.75 & 75 & 6.25 & 5.156 & 5.156 & 7.109 & 1.000 & 1.379 & -8.183 & 1.288 & -0.444 & FIM-7 \\
12.5 & 50 & 37.5 & 5.403 & 5.403 & 7.193 & 1.000 & 1.331 & -6.485 & 0.099 & -0.544 & FIM-3 \\
12.5 & 56.25 & 31.25 & 5.392 & 5.401 & 7.025 & 1.002 & 1.303 & -6.894 & 0.341 & -0.637 & FIM-3 \\
12.5 & 62.5 & 25 & 5.363 & 5.363 & 6.975 & 1.000 & 1.300 & -7.315 & 0.366 & -0.778 & FIM-7 \\
12.5 & 68.75 & 18.75 & 5.251 & 5.251 & 7.102 & 1.000 & 1.353 & -7.663 & 0.061 & -0.629 & FIM-7 \\
12.5 & 75 & 12.5 & 5.181 & 5.189 & 7.136 & 1.001 & 1.377 & -8.019 & 0.198 & -0.509 & FIM-7 \\
12.5 & 81.25 & 6.25 & 5.149 & 5.149 & 7.098 & 1.000 & 1.379 & -8.380 & 0.548 & -0.410 & FIM-3 \\
6.25 & 50 & 43.75 & 5.330 & 5.330 & 7.614 & 1.000 & 1.428 & -6.257 & 0.184 & -0.354 & FIM-3 \\
6.25 & 56.25 & 37.5 & 5.370 & 5.370 & 7.294 & 1.000 & 1.358 & -6.673 & 0.281 & -0.474 & FIM-3 \\
6.25 & 62.5 & 31.25 & 5.365 & 5.365 & 7.153 & 1.000 & 1.333 & -7.093 & 0.319 & -0.610 & FIM-7 \\
6.25 & 68.75 & 25 & 5.344 & 5.344 & 7.091 & 1.000 & 1.327 & -7.500 & 1.088 & -0.695 & FIM-7 \\
6.25 & 75 & 18.75 & 5.269 & 5.269 & 7.110 & 1.000 & 1.349 & -7.852 & 0.639 & -0.559 & FIM-7 \\
6.25 & 81.25 & 12.5 & 5.205 & 5.205 & 7.103 & 1.000 & 1.365 & -8.210 & 0.268 & -0.451 & FIM-7 \\
6.25 & 87.5 & 6.25 & 5.175 & 5.175 & 7.035 & 1.000 & 1.359 & -8.571 & 0.061 & -0.349 & FIM-7
\end{tabular}
\end{table}

\begin{table}[]
    \caption{Martensitic transition temperature $T_m$ for compositions of areas~I, II, and III with different concentrations of Ni, Mn, and Ga.}
    \label{table_T_m}
    \begin{tabular}{ccc|c|ccc|c|ccc|c}
    \multicolumn{4}{c|}{Area I} & \multicolumn{4}{c|}{Area II} & \multicolumn{4}{c}{Area III} \\
     \hline
    \multicolumn{3}{c|}{Concentration [at.\%]} & \multirow{2}{*}{$T_m$ [K]} & \multicolumn{3}{c|}{Concentration [at.\%]} & \multirow{2}{*}{$T_m$ [K]} & \multicolumn{3}{c|}{Concentration [at.\%]} & \multirow{2}{*}{$T_m$ [K]} \\
    Ni & Mn & Ga & & Ni & Mn & Ga & & Ni & Mn & Ga & 
 \\
 \hline
6.25  & 6.25  & 87.5  &     & 62.5  & 31.25 & 6.25  & 603 & 43.75 & 50    & 6.25  & 1163 \\
6.25  & 12.5  & 81.25 &     & 56.25 & 31.25 & 12.5  &     & 37.5  & 50    & 12.5  &      \\
6.25  & 18.75 & 75    &     & 56.25 & 37.5  & 6.25  &     & 37.5  & 56.25 & 6.25  &      \\
6.25  & 25    & 68.75 &     & 50    & 31.25 & 18.75 & 245 & 31.25 & 50    & 18.75 & 590  \\
12.5  & 6.25  & 81.25 &     & 50    & 37.5  & 12.5  & 397 & 31.25 & 56.25 & 12.5  &      \\
12.5  & 12.5  & 75    &     & 50    & 43.75 & 6.25  & 500 & 31.25 & 62.5  & 6.25  &      \\
12.5  & 18.75 & 68.75 &     & 43.75 & 31.25 & 25    & 105 & 25    & 50    & 25    & 338  \\
12.5  & 25    & 62.5  &     & 43.75 & 37.5  & 18.75 & 236 & 25    & 56.25 & 18.75 & 424  \\
18.75 & 6.25  & 75    & 332 & 43.75 & 43.75 & 12.5  & 389 & 25    & 62.5  & 12.5  & 686  \\
18.75 & 12.5  & 68.75 &     & 37.5  & 31.25 & 31.25 & 166 & 25    & 68.75 & 6.25  & 929  \\
18.75 & 18.75 & 62.5  &     & 37.5  & 37.5  & 25    & 154 & 18.75 & 50    & 31.25 & 176  \\
18.75 & 25    & 56.25 & 31  & 37.5  & 43.75 & 18.75 &     & 18.75 & 56.25 & 25    & 230  \\
25    & 6.25  & 68.75 & 515 & 31.25 & 31.25 & 37.5  &     & 18.75 & 62.5  & 18.75 & 403  \\
25    & 12.5  & 62.5  & 10  & 31.25 & 37.5  & 31.25 &     & 18.75 & 68.75 & 12.5  &      \\
25    & 18.75 & 56.25 &     & 31.25 & 43.75 & 25    & 155 & 18.75 & 75    & 6.25  & 1015 \\
25    & 25    & 50    &     & 25    & 31.25 & 43.75 & 189 & 12.5  & 50    & 37.5  & 221  \\
31.25 & 6.25  & 62.5  & 16  & 25    & 37.5  & 37.5  &     & 12.5  & 56.25 & 31.25 & 185  \\
31.25 & 12.5  & 56.25 &     & 25    & 43.75 & 31.25 &     & 12.5  & 62.5  & 25    & 356  \\
31.25 & 18.75 & 50    &     & 18.75 & 31.25 & 50    &     & 12.5  & 68.75 & 18.75 & 525  \\
31.25 & 25    & 43.75 & 26  & 18.75 & 37.5  & 43.75 &     & 12.5  & 75    & 12.5  &      \\
37.5  & 6.25  & 56.25 &     & 18.75 & 43.75 & 37.5  &     & 12.5  & 81.25 & 6.25  &      \\
37.5  & 12.5  & 50    &     & 12.5  & 31.25 & 56.25 &     & 6.25  & 50    & 43.75 & 386  \\
37.5  & 18.75 & 43.75 &     & 12.5  & 37.5  & 50    &     & 6.25  & 56.25 & 37.5  & 280  \\
37.5  & 25    & 37.5  &     & 12.5  & 43.75 & 43.75 &     & 6.25  & 62.5  & 31.25 & 453  \\
43.75 & 6.25  & 50    &     & 6.25  & 31.25 & 62.5  &     & 6.25  & 68.75 & 25    & 356  \\
43.75 & 12.5  & 43.75 &     & 6.25  & 37.5  & 56.25 &     & 6.25  & 75    & 18.75 & 527  \\
43.75 & 18.75 & 37.5  &     & 6.25  & 43.75 & 50    & 681 & 6.25  & 81.25 & 12.5  &      \\
43.75 & 25    & 31.25 &     &       &       &       &     & 6.25  & 87.5  & 6.25  &      \\
50    & 6.25  & 43.75 &     &       &       &       &     &       &       &       &      \\
50    & 12.5  & 37.5  &     &       &       &       &     &       &       &       &      \\
50    & 18.75 & 31.25 &     &       &       &       &     &       &       &       &      \\
50    & 25    & 25    & 96  &       &       &       &     &       &       &       &      \\
56.25 & 6.25  & 37.5  &     &       &       &       &     &       &       &       &      \\
56.25 & 12.5  & 31.25 & 23  &       &       &       &     &       &       &       &      \\
56.25 & 18.75 & 25    & 127 &       &       &       &     &       &       &       &      \\
56.25 & 25    & 18.75 & 228 &       &       &       &     &       &       &       &      \\
62.5  & 6.25  & 31.25 &     &       &       &       &     &       &       &       &      \\
62.5  & 12.5  & 25    &     &       &       &       &     &       &       &       &      \\
62.5  & 18.75 & 18.75 & 186 &       &       &       &     &       &       &       &      \\
62.5  & 25    & 12.5  & 441 &       &       &       &     &       &       &       &      \\
68.75 & 6.25  & 25    & 277 &       &       &       &     &       &       &       &      \\
68.75 & 12.5  & 18.75 & 257 &       &       &       &     &       &       &       &      \\
68.75 & 18.75 & 12.5  & 465 &       &       &       &     &       &       &       &      \\
68.75 & 25    & 6.25  & 598 &       &       &       &     &       &       &       &      \\
75    & 6.25  & 18.75 & 398 &       &       &       &     &       &       &       &      \\
75    & 12.5  & 12.5  & 547 &       &       &       &     &       &       &       &      \\
75    & 18.75 & 6.25  &     &       &       &       &     &       &       &       &      \\
81.25 & 6.25  & 12.5  & 703 &       &       &       &     &       &       &       &      \\
81.25 & 12.5  & 6.25  &     &       &       &       &     &       &       &       &      \\
87.5  & 6.25  & 6.25  & 527 &       &       &       &     &       &       &       &   
\end{tabular}%
\end{table}

\newpage
\bibliography{main}